\documentclass[12pt]{article}
\usepackage{psfrag}
\usepackage{amsmath}
\usepackage{amsthm}
\usepackage{amssymb}
\usepackage{epsfig}
\usepackage{euscript}
\usepackage{array}
\usepackage{bbold}
\usepackage{cancel}
\usepackage{mathtools}
\usepackage{empheq}
\usepackage{graphicx}
\usepackage{subfigure}
\usepackage{upgreek}
\usepackage{epstopdf}
\usepackage{booktabs}
\usepackage{mathrsfs}
\usepackage{amsbsy}

\usepackage{hyperref}
\usepackage{verbatim}

\theoremstyle{definition}

\theoremstyle{remark}

\newcounter{multieqs}

\newcommand{\be}{\begin{equation}}
\newcommand{\ee}{\end{equation}}
\newcommand{\eq}[1]{(\ref{#1})}
\newcommand{\bit}{\begin{itemize}}  \newcommand{\eit}{\end{itemize}}
\newcommand{\ben}{\begin{enumerate}}  \newcommand{\een}{\end{enumerate}}

\newcommand{\ket}[1]{|#1 \rangle}

\newcommand{\bm}[1]{\mbox{\boldmath $#1$}}
\newcommand{\rf}[1]{(\ref{#1})}

\def\bd{\begin{document}}
\def\ed{\end{document}}
\def\bea{\begin{eqnarray}}
\def\eea{\end{eqnarray}}
\let\bm=\bibitem

\def\la{\langle}
\def\ra{\rangle}

\def\npb#1#2#3{Nucl. Phys. {\bf{B#1}} #3 (#2)}
\def\plb#1#2#3{Phys. Lett. {\bf{#1B}} #3 (#2)}
\def\prl#1#2#3{Phys. Rev. Lett. {\bf{#1}} #3 (#2)}
\def\prd#1#2#3{Phys. Rev. {D bf{#1}} #3 (#2)}
\def\cmp#1#2#3{Comm. Math. Phys. {\bf{#1}} #3 (#2)}
\def\cqg#1#2#3{Class. Quantum Grav. {\bf{#1}} #3 (#2)}
\def\nppsa#1#2#3{Nucl. Phys. B (Proc. Suppl.) {\bf{#1A}}#3 (#2)}
\def\ap#1#2#3{Ann. of Phys. {\bf{#1}} #3 (#2)}
\def\ijmp#1#2#3{Int. J. Mod. Phys. {\bf{A#1}} #3 (#2)}
\def\rmp#1#2#3{Rev. Mod. Phys. {\bf{#1}} #3 (#2)}
\def\mpla#1#2#3{Mod. Phys. Lett. {\bf A#1} #3 (#2)}
\def\jhep#1#2#3{J. High Energy Phys. {\bf #1} #3 (#2)}
\def\atmp#1#2#3{Adv. Theor. Math. Phys. {\bf #1} #3 (#2)}

\def\N{{\cal N}}
\def\sst{\scriptscriptstyle}
\def\thetabar{\bar\theta}
\def\Tr{{\rm Tr}}
\def\one{\mbox{1 \kern-.59em {\rm l}}}

\def\a{\alpha}      \def\da{{\dot\alpha}}  \def\dA{{\dot A}}
\def\b{\beta}       \def\db{{\dot\beta}}
\def\g{\gamma}  \def\G{\Gamma}  \def\dc{{\dot\gamma}}
\def\d{\delta}  \def\D{\Delta}  \def\ddt{\dot\delta}
\def\e{\epsilon}
\def\ve{\varepsilon}
\def\uve{\upvarepsilon}
\def\f{\phi}    \def\F{\Phi}    \def\vvf{\f}
\def\vphi{\varphi}
\def\h{\eta}
\def\k{\kappa}
\def\l{\lambda} \def\L{\Lambda}
\def\m{\mu} \def\n{\nu}
\def\o{\omega}
\def\p{\pi} \def\P{\Pi}
\def\r{\rho}
\def\s{\sigma}  \def\S{\Sigma}
\def\t{\tau}
\def\th{\theta} \def\Th{\Theta} \def\vth{\vartheta}
\def\X{\Xeta}
\def\z{\zeta}

\def\na{\nabla}
\def\cA{{\mathscr A}} \def\cB{{\cal B}} \def\cC{{\cal C}}
\def\cD{{\cal D}} \def\cE{{\cal E}} \def\cF{{\cal F}}
\def\cG{{\cal G}} \def\cH{{\cal H}} \def\cI{{\cal I}}
\def\cJ{{\mathscr J}} \def\cK{{\cal K}} \def\cL{{\cal L}}
\def\cM{{\cal M}} \def\cN{{\cal N}} \def\cO{{\cal O}}
\def\cP{{\cal P}} \def\cQ{{\cal Q}} \def\cR{{\cal R}}
\def\cS{{\cal S}} \def\cT{{\cal T}} \def\cU{{\cal U}}
\def\cV{{\cal V}} \def\cW{{\cal W}} \def\cX{{\cal X}}
\def\cY{{\cal Y}} \def\cZ{{\cal Z}}

\def\ua{\underline{\alpha}}
\def\uc{\underline{\phantom{\alpha}}\!\!\!\gamma}
\def\um{\underline{\mu}}
\def\ud{\underline\delta}
\def\ue{\underline\epsilon}
\def\una{\underline a}\def\unA{\underline A}
\def\unb{\underline b}\def\unB{\underline B}
\def\unc{\underline c}\def\unC{\underline C}
\def\und{\underline d}\def\unD{\underline D}
\def\une{\underline e}\def\unE{\underline E}
\def\unf{\underline{\phantom{e}}\!\!\!\! f}\def\unF{\underline F}
\def\unm{\underline m}\def\unM{{\underline M}}
\def\unn{\underline n}\def\unN{{\underline N}}
\def\unp{\underline{\phantom{a}}\!\!\! p}\def\unP{\underline P}
\def\unq{\underline{\phantom{a}}\!\!\! q}
\def\unQ{\underline{\phantom{A}}\!\!\!\! Q}
\def\unH{\underline{H}}

\def\As {{A \hspace{-6.4pt} \slash}\;}
\def\bs {{b \hspace{-6.4pt} \slash}\;}
\def\Ds {{D \hspace{-6.4pt} \slash}\;}
\def\Gts {{\Gt \hspace{-6.4pt} \slash}\;}
\def\ds {{\del \hspace{-6.4pt} \slash}\;}
\def\ss {{\s \hspace{-6.4pt} \slash}\;}
\def\ks {{ k \hspace{-6.4pt} \slash}\;}
\def\ps {{p \hspace{-6.4pt} \slash}\;}
\def\xs {{x \hspace{-6.4pt} \slash}\;}
\def\pas {{{p_1} \hspace{-6.4pt} \slash}\;}
\def\pbs {{{p_2} \hspace{-6.4pt} \slash}\;}
\def\cFs {{{\cal F} \hspace{-6.4pt} \slash}\;}
\def\Dss {{D \hspace{-7.5pt} \slash}\;}
\def\dss {{\del \hspace{-7.0pt} \slash}\;}

\def\Ah{{\hat{A}}}
\def\Dh{{\hat{D}}}
\def\Gh{{\hat{G}}}
\def\Fh{{\hat{F}}}
\def\Ih{{\hat{I}}}
\def\Jh{{\hat{J}}}
\def\Kh{{\hat{K}}}
\def\Lh{{\hat{L}}}
\def\Ph{{\hat{P}}}
\def\Rh{{\hat{R}}}
\def\Vh{{\hat{V}}}
\def\Xh{{\hat{X}}}

\def\ah{{\hat{\a}}}
\def\bh{{\hat{\b}}}
\def\gh{{\hat{\g}}}
\def\dh{{\hat{\d}}}
\def\rh{{\hat{\r}}}
\def\hh{\hat{h}}
\def\uh{\hat{u}}
\def\xh{\hat{x}}
\def\yh{\hat{y}}
\def\ph{\hat{p}}
\def\xih{\hat{\xi}}
\def\chih{\hat{\chi}}
\def\Psih{\hat{\Psi}}
\def\phih{\hat{\phi}}

\def\psit{\tilde{\psi}}
\def\Psit{\tilde{\Psi}}
\def\Psibt{\tilde{\bar{Psi}}}

\def\st{\tilde{\sigma}}

\def\delt{\tilde{\delta}}
\def\Phit{\tilde{\Phi}}
\def\Phitb{\overline{\tilde{Phi}}}
\def\tht{\tilde{\th}}
\def\lt{\tilde{\l}}
\def\chit{\tilde{\chi}}
\def\phit{\tilde{\phi}}

\def\At{\tilde{A}}
\def\Bt{\tilde{B}}
\def\Ct{\tilde{C}}
\def\Dt{\tilde{D}}
\def\Et{\tilde{E}}
\def\Ft{\tilde{F}}
\def\Gt{\tilde{G}}
\def\Ht{\tilde{H}}
\def\It{\tilde{I}}
\def\Jt{\tilde{J}}
\def\Qt{\tilde{Q}}
\def\Rt{\tilde{R}}
\def\Mt{\tilde{M }}
\def\Nt{\tilde{N}}
\def\St{\tilde{S}}
\def\Vt{\tilde{V}}
\def\Xt{\tilde{X}}
\def\at{\tilde{a}}
\def\ct{\tilde{c}}
\def\dt{\tilde{d}}
\def\htt{\tilde{h}}
\def\ft{\tilde{f}}
\def\gt{\tilde{g}}
\def\pt{\tilde{p}}
\def\qt{\tilde{q}}
\def\vt{\tilde{v}}
\def\nt{\tilde{n}}
\def\ut{\tilde{u}}
\def\wt{\tilde{w}}
\def\zt{\tilde{z}}
\def\xt{\tilde{x}}
\def\yt{\tilde{y}}
\def\Psit{\tilde{\Psi}}
\def\vphit{\tilde{\varphi}}
\def\tD{\tilde{\D}}

\def\eb{\bar{\epsilon}}
\def\delb{\bar{\partial}}
\def\thb{\bar{\theta}}
\def\mub{\bar{\mu}}
\def\lamb{\bar{\l}}
\def\psib{\bar{\psi}}
\def\sb{\bar{\sigma}}
\def\xib{\bar{\xi}}
\def\chib{\bar{\chi}}

\def\Psib{\bar{\Psi}}
\def\Phib{\bar{\Phi}}
\def\Lamb{\bar{\Lambda}}
\def\Sb{{\overline \Sigma}}
\def\cb{\bar{c}}
\def\hb{\bar{h}}
\def\qb{\bar{q}}
\def\wb{\bar{w}}
\def\ub{\bar{u}}
\def\zb{{\bar{z}}}
\def\Hb{\bar{H}}
\def\Qb{{\bar Q}}
\def\Omegab{\overline{\Omega}}
\def\ob{\overline{\omega}}

\def\Ab{{\overline A}} \def\Bb{{\overline B}} \def\Cb{{\overline C}}
\def\Db{{\overline D}} \def\Eb{{\overline E}} \def\Fb{{\overline F}}
\def\Gb{{\overline G}}
\def\Ib{{\overline I}}
\def\Jb{{\overline J}} \def\Kb{{\overline K}} \def\Lb{{\overline L}}
\def\Mb{{\overline M}} \def\Nb{{\overline N}} \def\Ob{{\overline O}}
\def\Pb{{\overline P}}  \def\Rb{{\overline R}}
 \def\Tb{{\overline T}} \def\Ub{{\overline U}}
\def\Vb{{\overline V}} \def\Wb{{\overline W}} \def\Xb{{\overline X}}
\def\Yb{{\overline Y}} \def\Zb{{\overline Z}}

\def\fb{{\overline f}}
\def\gb{{\overline g}}
\def\mb{{\overline m}}
\def\lb{{\overline l}}
\def\yb{{\overline y}}

\def\ldel{{\overleftarrow{\del}}}
\def\rdel{{\overrightarrow{\del}}}
\def\ldeldel{{\overleftarrow{\del^2}}}
\def\rdeldel{{\overrightarrow{\del^2}}}
\def\ldelb{{\overleftarrow{\bar{\del}}}}
\def\rdelb{{\overrightarrow{\bar{\del}}}}
\def\ba{{\bf a}}
\def\bk{{\bf k}}
\def\bl{{\bf l}}
\def\bp{{\bf p}}
\def\bq{{\bf q}}
\def\br{{\bf r}}
\def\bt{{\bf t}}
\def\bu{{\bf u}}
\def\bv{{\bf v}}
\def\bx{{\bf x}}
\def\by{{\bf y}}
\def\bA{{\bf A}}
\def\bB{{\bf B}}
\def\bR{{\bf R}}
\def\bV{{\bf V}}

\def\bz{{\boldsymbol{\zeta}}}

\def\bone{{\bf 1}}

\def\va{{\vec a}}
\def\vk{{\vec k}}
\def\vp{{\vec p}}
\def\vq{{\vec q}}
\def\vx{{\vec x}}
\def\vy{{\vec y}}
\def\vu{{\vec u}}
\def\vv{{\vec v}}
\def \vH{{\vec H}}
\def \vg{{\vec g}}

\def\vs{{\vec \sigma}}
\def\vtau{{\vec \tau}}

\newcommand{\ov}[1]{\overrightarrow{#1}}

\def\frA{\mathfrak{A}}
\def\frB{\mathfrak{B}}
\def\frC{\mathfrak{C}}
\def\frD{\mathfrak{D}}
\def\frE{\mathfrak{E}}
\def\frF{\mathfrak{F}}
\def\frG{\mathfrak{G}}
\def\frH{\mathfrak{H}}
\def\frM{\mathfrak{M}}
\def\frN{\mathfrak{N}}
\def\frR{\mathfrak{R}}
\def\frW{\mathfrak{W}}

\def\fra{\mathfrak{a}}
\def\frb{\mathfrak{b}}
\def\frf{\mathfrak{f}}
\def\frg{\mathfrak{g}}
\def\frh{\mathfrak{h}}
\def\frl{\mathfrak{l}}
\def\frs{\mathfrak{s}}
\def\fri{\mathfrak{i}}
\def\frj{\mathfrak{j}}

\def\ma{\mathfrak{a}}
\def\mg{\mathfrak{g}}
\def\mh{\mathfrak{h}}
\def\mR{\mathfrak{R}}
\def\mN{\mathfrak{N}}

\newcommand{\nn}{{\nonumber}}

\def\d{\delta}\def\D{\Delta}\def\ddt{\dot\delta}

\def\pa{\partial} \def\del{\partial}
\def\xx{\times}
\def\uno{\mbox{1 \kern-.59em {\rm l}}}

\def\trp{^{\top}}
\def\inv{^{-1}}
\def\dag{\dagger}
\def\pr{^{\prime}}

\def\rar{\rightarrow}
\def\lar{\leftarrow}
\def\lrar{\leftrightarrow}

\newcommand{\0}{\,\!}      %this is just NOTHING!
\def\one{1\!\!1\,\,}
\def\im{\imath}
\def\jm{\jmath}

\newcommand{\tr}{\mbox{tr}}
\newcommand{\slsh}[1]{/ \!\!\!\! #1}

\def\vac{|0\rangle}
\def\lvac{\langle 0|}

\def\hlf{\frac{1}{2}}
\def\ove#1{\frac{1}{#1}}
\newcommand{\hot}[1]{\frac{#1}{2}}

\def\Box{\square}
\def\CC {\mathbb{C}}
\def\FF {\mathbb{F}}
\def\RR{\mathbb{R}}
\def\NN{\mathbb{N}}
\def\ZZ{\mathbb{Z}}
\def\bb#1{{\bf #1}}
\def\bcomment#1{}
\def\bfhat#1{{\bf \hat{#1}}}
\def\VEV#1{\left\langle #1\right\rangle}

\newcommand{\ex}[1]{{\rm e}^{#1}} \def\ii{{\rm i}}

\newcommand{\lrbrk}[1]{\left(#1\right)}
\newcommand{\lrsbrk}[1]{\left[#1\right]}
\newcommand{\sfrac}[2]{{\textstyle\frac{#1}{#2}}}

\def\stw{{\sqrt{2}}}

\def\rf {{\rm f}}
\def\ri {{\rm i}}
\def\rj {{\rm j}}
\def\rn {{\rm n}}
\def\rk {{\rm k}}
\def\rl {{\rm l}}
\def\rr {{\rm r}}
\def\rs {{\scriptscriptstyle \rm S}}
\def\rt {{\scriptscriptstyle \rm T}}

\def\rQ {{\scriptscriptstyle \rm \cQ}}
\def\rR {{\scriptscriptstyle \rm \cR}}

\def\cQb{{\cal \Qb}}
\def\cRb{{\cal \Rb}}
\def\cWb{{\cal \Wb}}

\def\fd {{\rm N}}
\def\afd {{\overline{\rm N}}}

\def \II {I\hspace{-.1em}I\hspace{.1em}}
\def \IIA {\mbox{\II A\hspace{.2em}}}
\def \IIB {\mbox{\II B\hspace{.2em}}}
\def \gs {g^s}
\def \ls {\lambda^s}

\def \I {{\cal I}}
\def \qs {q\hspace{-.53em}/\hspace{.15em}}
\def \ks {k\hspace{-.53em}/\hspace{.15em}}
\def \YM {{\mbox{\tiny YM}}}
\def \gym {g_{\YM}}

\def \Lc {\L_c}
\def\IR{\relax{\rm I\kern-.18em R}}
\def \id {{\bf 1}}

\def\cci{\ell}
\def\ccj{\ell'}

\def\bbq{\pmb{q}}

\begin{document}
\begin{titlepage}
\begin{flushright}
\hfill{ NCTS-TH/1609} 
\end{flushright}
\hfill
% \vskip 0.4in

 \begin{center}

%%%%%%%%%%%%%%%%%%%%%%%%%%%%%%%%%%%%%%%%%%%%%%%%%%
{\Large \bf Adiabatic Regularization for Gauge Field 
\\
and the Conformal Anomaly
}\\[10mm]
%%%%%%%%%%%%%%%%%%%%%%%%%%%%%%%%%%%%%%%%%%%%%%%%%%

{\bf Chong-Sun Chu${}^{1,2}$,  Yoji Koyama${}^1$}

{\itshape ${}^1$ Physics Division, National Center for Theoretical
  Sciences, \\
 National Tsing-Hua University, Hsinchu, 30013, Taiwan}\\[1mm]
{\itshape ${}^2$ Department of Physics, National Tsing-Hua
  University,  Hsinchu 30013, Taiwan}

%d1
{\small \sffamily
cschu@phys.nthu.edu.tw~, koyama811@cts.nthu.edu.tw
\\
}
\end{center}

\date{\today}

\begin{abstract}

We construct 
%c2 and provide 
the adiabatic regularization method for a $U(1)$ gauge field in a 
conformally flat spacetime by quantizing in the canonical formalism 
the gauge fixed $U(1)$ theory with mass
terms for the gauge fields and the ghost fields.
We show that  the adiabatic expansion for the mode functions 
and the adiabatic
vacuum can be defined in a similar way  using  
WKB-type solutions as the scalar fields.  
As an application of the adiabatic method, 
we compute the trace of the
energy momentum tensor and
reproduces the known result for the conformal
anomaly obtained by the other regularization methods. The availability
of the adiabatic expansion scheme for gauge field allows one
to study the renormalization of the de-Sitter space maximal
superconformal Yang-Mills theory using the adiabatic regularization
method.

\end{abstract}

\end{titlepage}
\newpage

% \tableofcontents

%%%%%%%%%%%%%%%%%%%%%%%%%%%%%%%%%%%%%%%%%%%%%%%%%%%%%%%%%%%%%%%%%%%%%%%
\section{Introduction}
%%%%%%%%%%%%%%%%%%%%%%%%%%%%%%%%%%%%%%%%%%%%%%%%%%%%%%%%%%%%%%%%%%%%%%%%%%

The study of the dynamics of quantum field theory in curved spacetime
is not only relevant for the understanding of a number of
important physical problems such as inflation or  Hawking radiation, to
name a few, it is also rather challenging. See, for example, the books
\cite{Birrell:1982ix,Calzetta:2008iqa,Parker:2009uva} for a general exposition. 
One of the challenges is
the determination of the vacuum. In fact, 
as time is not a diffeomorphic invariant concept, 
neither is the vacuum. The observer dependent nature of the vacuum is
therefore intrinsic to QFT in curved spacetime. Even after one fixes
a choice of time, the vacuum in perturbation theory 
is generally still not unique. For metric with 
isometries, it is often preferred to choose the vacuum to respect the
symmetries. But still one may not be able to get a unique one. 
For example for the de-Sitter metric, the alpha-vacua give
a one parameter family of vacua which are invariant under the
de-Sitter isometries, and one is able to single out the Bunch-Davies
vacuum only if the Hadamard property is also imposed. 

After one decided on the vacuum, one could then proceed to study various
quantum properties of the system using traditional tools of quantum
field theory in flat spacetime. However extra care must be exercised
to take into account of the effects of particle creation which is a
simple consequence of the fact that, in general for a time dependent
background, a vacuum at time $t$ may not be a vacuum
anymore at a different time $t'$. As a result, instead of S-matrix,
it is more sensible to consider correlation functions of operators
%c2 ref added
\cite{Chou:1984es,Calzetta:1986ey,Jordan:1986ug,Weinberg:2005vy}
for  QFT in curved spacetime. 

Historically, the conformal 
%c2
(Weyl) anomaly
of the energy momentum tensor was one of the first quantities studied
and computed for a QFT in curved spacetime.
And various methods have been developed to
regularize the UV divergence found in the energy momentum tensor.
These includes, for example, the Dewitt-Schwinger geodesic 
point splitting method \cite{dewitt}, zeta-function
regularization \cite{Dowker:1976zf,Hawking:1976ja} and the adiabatic
regularization method \cite{Parker:1974qw}.

Adiabatic regularization \cite{Parker:1974qw,Fulling:1974pu} is a
useful and simple method to obtain physically meaningful renormalized
results from the formally UV divergent quantities, e.g. the vacuum
expectation value of the energy momentum tensor, in an expanding
universe such as a conformally flat spacetime.  The most studied example is the
adiabatic regularization for scalar fields
\cite{Parker:1974qw,Fulling:1974pu,Bunch:1978gb,Bunch:1980vc} (see
also \cite{Haro:2010mx} for a recent review and references
therein). Recently the adiabatic regularization for fermion has been
established \cite{Landete:2013lpa,delRio:2014cha}.  On the other hand,
as far as we know, adiabatic regularization for gauge
fields has never been considered in the literature. One of the
motivation of this paper is to fill this gap. 

At the first
glance, one may think that 
the adiabatic regularization for gauge field is
rather straightforward since the theory of a massless gauge field in a 
4-dimensional  flat spacetime is conformally invariant. 
One may then infer that the mode function of a gauge field $A_\mu$ can be
collectively written in the same form, up to some overall scaling
factor, as that of a massless
conformally coupled  scalar field, and thus
the adiabatic expansion for the gauge field can be performed exactly in the
same way as that for a massless conformally coupled scalar field.  
This is actually wrong and one would get the wrong result for the
conformal anomaly. 
The reason for the mistake is that 
one has missed  a very important nontrivial issue related to the 
gauge fixing of the theory, the latter of which is
essential to the setting up of the perturbation theory. 

 The conformal anomaly for gauge field has been obtained using other
regularization schemes before
%c1
\cite{Capper:1974ic,Deser:1976yx,Christensen:1978md,
Brown:1977pq,Duff:1977ay,
Duff:1993wm,Endo:1984sz,Toms:2014tia}
%c2
and is given by
%yoji1006   
\footnote{General form of conformal anomaly in arbitrary dimensions was
obtained by momentum space calculation in \cite{Deser:1993yx}.}
\be \label{other-v}
\langle T^\m_\m \rangle =
\frac{1}{2880\pi^2}
\left[
62\Big(R_{\mu\nu}R^{\mu\nu}-\frac13R^2\Big)
+ d\, \Box R
\right],
\ee
where it is known that there is a
discrepancy  in the 
%c2
coefficient $d$ of $\Box R$ among different schemes: dimensional
regularization  gives  $12$ \cite{Brown:1977pq,Duff:1977ay},
while the 
%c2 other regularization methods 
DeWitt-Schwinger point-splitting expansion gives $-18$
\cite{Dowker:1976zf}. 
%c2
In fact it is well understood that this term is regularization
dependent since it can be expressed  as the variation of a local action
\cite{Brown:1977pq,Duff:1977ay}: 
\bea
\sqrt{-g} \Box R
=
\frac{1}{6 }g^{\mu\nu}\frac{\delta}{\delta g^{\mu\nu}}
\int d^4x \sqrt{-g} R^2
\label{boxr}
\eea
and so the value of $d$ can be shifted to any arbitrary value by using
an appropriate counter term. 
%c2
Regularization dependence of the 
$\Box R$ term has also been discussed recently 
in \cite{Vieira:2015oka}. 
%c2 
It has also been pointed out 
\cite{Endo:1984sz,Toms:2014tia} that
the coefficient of $\Box R$, 
%c2 
at least  in the DeWitt-Schwinger  regularization scheme,
is gauge dependent. 
%yoji1006
%c2 Regularization dependence and intrinsic ambiguity of the
%coefficient of $\Box R$ is recently discussed in detail 
% in \cite{Vieira:2015oka}. 
%
%c2 
We are interested not only in computing the conformal anomaly of gauge
field using the adiabatic method, but also to 
%c2
compare our results,
especially the gauge dependence 
of the $\Box R$ term, with those obtained in the other regularization methods.
%c2 find out if and which of the above mentioned results will be
%obtained. 
This is another motivation of this paper.

Recently the 
$\cN=4$ superconformal Yang-Mills theory on de-Sitter space
\cite{Anous:2014lia} has been introduced and it has been proposed
\cite{Chu:2016uwi} to be the holographic dual of the type IIB string
theory on $AdS_5 \times S^5$ background 
with certain boundary conditions. Furthermore, the holographic
duality suggests that the de-Sitter space 
superconformal Yang-Mills theory has 
a number of rather interesting quantum 
properties similar to that of the maximal superconformal Yang-Mills theory on
flat spacetime. To check,  a consistent framework of
evaluating the quantum loop contributions in the
conformally flat spacetime is necessary. 
Compare to the other regularization schemes, 
the adiabatic regularization scheme is practical and particular useful
for perturbative
quantum field theory computation in a conformally flat metric as
it has taken full advantage of the homogeneity of the metric. As a
result,   the
mode expansion of the field can be greatly 
simplified and one simply obtain  an oscillator with 
time dependent
frequency,  whose solution can be obtained via an adiabatic expansion
in terms of slowness of the temporal change of the metric. However, 
while the adiabatic
regularization schemes for scalar field  and fermion field
are available, the adiabatic scheme for gauge field has not
been constructed before.
The main motivation of this work is indeed to develop such a scheme
for the gauge field so that one has available a practical
and complete framework in which one can use to handle 
the UV divergences and study the renormalization of the theory.

The next section is devoted to a brief review of the adiabatic
regularizations for a scalar field and for a Dirac fermion. In section
3, we consider the adiabatic expansion for $U(1)$ gauge theory. In
section 4, we compute the conformal anomaly for the $U(1)$ gauge
theory in the adiabatic regularization. We summarize our result in
section 5.

Our convention of the Minkowski metric is
$\eta_{\mu\nu}=\text{diag}(-1,1,1,1)$ 
and the Riemann and Ricci tensors are given by 
\be
R^{\rho}_{\ \sigma\mu\nu}
=\partial_\mu\Gamma^{\rho}_{\ \nu\sigma}-\partial_{\nu}\Gamma^{\rho}_{\ \mu\sigma}
+\Gamma^{\rho}_{\ \alpha\mu}\Gamma^{\alpha}_{\ \sigma\nu}
-\Gamma^{\rho}_{\ \alpha\nu}\Gamma^{\alpha}_{\ \sigma\mu},
\qquad 
R_{\sigma\nu}=R^{\rho}_{\ \sigma\rho\nu}.
\ee 
Note that sign convention on 
%c1 
the signature of the metric affects the 
%c1 signature 
signs of the d'Alembertian operator, 
and the convention of the Riemann and Ricci tensors
affects the 
%c1 signature 
sign of the scalar curvature $R$. In the conformal
anomaly the 
%c1 signature 
overall sign of the  $\Box R$ term is thus convention
dependent. For example, $\Box R$ in our convention has 
%c1
the same sign as those in \cite{Duff:1977ay,Duff:1993wm} 
and opposite sign as those in \cite{Dowker:1976zf,Endo:1984sz}.

%%%%%%%%%%%%%%%%%%%%%%%%%%%%%%%%%%%%%%%%%%%%%%%%%%%%%%%%%%%%%%%%%%%
\section{Adiabatic regularizations for scalar field and Dirac fermion
  in conformally flat spacetime}
%%%%%%%%%%%%%%%%%%%%%%%%%%%%%%%%%%%%%%%%%%%%%%%%%%%%%%%%%%%%%%%%%%

In order to see the basic strategy of the adiabatic method, 
in this section we give a brief review of the
adiabatic expansions and regularizations for a scalar field and for a
Dirac fermion in a conformally flat spacetime.

%%%%%%%%%%%%%%%%%%%%%%%%%%%%%%%%%%%%%%%%%%%%%%%%%%%%%%%%%%%%%%%%%%%%%%%%%%%%%%%%%%%%%%%%
\subsection{Conformally coupled scalar field}
%%%%%%%%%%%%%%%%%%%%%%%%%%%%%%%%%%%%%%%%%%%%%%%%%%%%%%%%%%%%%%%%%%%%%%%%%%%%%%%%%%%%%%%%

We consider a conformally flat spacetime with metric 
\bea
%c1 ds^2=C(\tau)(-d\tau^2+\delta_{ij}dx^idx^j),
g_{\m\n} = C(\t) \eta_{\m\n}, \qquad x^\m = (\t, x^i), \quad i=1,2,3,
\label{metric}
\eea
where $C(\tau)\equiv a(\tau)^2$ and $a(\tau)$ is the cosmological scale
factor. 
In
order to perform the adiabatic expansion, we need to introduce a mass
$m$ to the scalar field and take a zero mass limit in
the end of calculation. This mass will play a role in capturing the
effect of background gravitational field.  The action is given by
\bea
S
=
\int d^4 x \sqrt{-g}
\left(
-\frac12 g^{\mu\nu}\partial_\mu\phi\partial_\nu\phi
-\frac12\Big(m^2 + \frac{R}{6}\Big) \phi^2
\right),
\eea 
with the field equation
\bea
g^{\mu\nu}\nabla_\mu\nabla_\nu \phi - \Big(m^2+\frac{R}{6}\Big)\phi=0,
%(-\partial_0^2+\delta^{ij}\partial_i\partial_j-m^2C)(C^{\frac12}\phi)=0.
\label{feqscalar}
\eea
where $\nabla_{\mu}$ is the covariant derivative associated with 
the background metric.
This gives the energy momentum tensor 
\begin{align}
T_{\mu\nu}
&=
\frac{-2}{\sqrt{-g}}\frac{\delta S}{\delta g^{\mu\nu}}
=
\frac23\partial_\mu\phi \partial_\nu\phi
-\frac{1}{6}g_{\mu\nu}
g^{\rho\sigma}\partial_\rho\phi\partial_{\sigma}\phi 
-\frac12 g_{\mu\nu} m^2\phi^2
-\frac13 \phi \nabla_\mu \nabla_\nu \phi
\notag\\
&\hspace{3cm}
+\frac{1}{3}g_{\mu\nu}\phi g^{\rho\sigma}\nabla_\rho\nabla_\sigma \phi
+\frac16 \Big(R_{\mu\nu}-\frac12 g_{\mu\nu} R\Big) \phi^2,
\end{align}
and the trace 
\be
T^\mu_{\ \mu} =
-m^2\phi^2,
\label{ctracescalar}
\ee
where the field equation \eqref{feqscalar} 
has been used to obtain \eqref{ctracescalar}.
The field equation \eq{feqscalar} 
can be solved with the Fourier expansion
\bea
\phi(x)
=
\frac{1}{\sqrt{C}}\int\frac{d^3k}{(2\pi)^3}
\Big(
a_{\vec k}\varphi(\tau,k)e^{i{\vec k}\cdot{\vec x}}+\text{h.c}
\Big),
\eea
where $k=|{\vec k}|$ and the 
mode function $\varphi(\tau,k)$ satisfies a second order differential
equation which is precisely that of 
a harmonic oscillator with a time dependent frequency
\bea
(\partial_0^2+\omega^2)\vphi(\tau,k)=0,
\qquad
\omega^2=k^2+m^2C.
\label{meqscalar}
\eea
The operators $a_{\vk}$ satisfies the commutation relation 
of creation and annihilation
operators 
\be
[a_{\vec k},a^{\dagger}_{{\vec k}'}]
=(2\pi)^3\delta^{(3)}({\vec k}-{\vec k}'), \qquad 
[a_{\vec    k},a_{{\vec k}'}] =0
\ee
iff the mode function $\vphi(\t,k)$ 
satisfies  the normalization condition
\be
\vphi(\tau,k)\partial_0\vphi^\ast(\tau,k)
-\partial_0\vphi(\tau,k)\vphi^\ast(\tau,k)=i.
\label{normalizationscalar}
\ee
The vacuum of the theory is defined to be a state annihilated by the 
operators $a_\vk$.
However this depends on the choice of the mode function as different
choices of the mode functions determine different set of annihilation
operators $a_\vk$, and hence the vacuum states. In general solving 
analytically \eq{meqscalar} is impossible. An useful observation 
is that  the normalization condition
\eq{normalizationscalar} can be conveniently solved by
\cite{Parker:1974qw}
\be \label{gen-vphi}
\vphi(\tau,k)
= \frac{1}{\sqrt{2W(\t)}}
\left(\a e^{-i\int^\tau W(\tau') d\tau'} + \b 
e^{i\int^\tau W(\tau') d\tau'} \right), 
\ee
where $\a,\b$ are constant coefficients satisfying
\be
|\a|^2 - |\b|^2 =1
\ee
and $W(\t)$ is an arbitrary function. 
The differential equation \eq{meqscalar} for $\vphi$ becomes the
differential equation for $W$:
\bea
W^2
=\omega^2
- \left(\frac{W''}{2W}-\frac{3 (W')^2}{4 W^2}\right),
\label{wkbequationscalar}
\eea
where prime denotes differential with respect to the conformal time, 
$'\equiv \partial_0=\partial/ \partial \tau$. 
This equation is complicated and, again, impossible to solve
analytically in general. However if one consider the background to 
be slowly changing and parametrize the time variation by a small
parameter $\e\ll 1$: $\del_0 \to \e \del_0$,  then the equation 
\eq{wkbequationscalar} can be solved iteratively to give rises to an
expansion  in powers of time derivatives
\bea
W
= W_{(0)}+ \epsilon^2W_{(2)}+ \epsilon^4W_{(4)}+\cdots.
\label{adiabaticscalar}
\eea
Here $W_{(n)}$ contains $n$ orders of time derivatives.
The expansion \eq{adiabaticscalar} is a WKB-type expansion and 
defines the adiabatic
expansion of the mode function of the scalar field, with 
$n$ being called the order of the adiabatic expansion.
The first few terms of the expansion are
\begin{align}
W_{(0)}
&=
\omega, \label{wkbzerothscalar} \\
W_{(2)}
&=
\frac{3}{8}\frac{(\omega')^2}{\omega^3}
-\frac{1}{4}\frac{\omega''}{\omega^2},
\\
W_{(4)}
&=
-\frac{297}{128}\frac{(\omega')^4}{\omega^7}
+\frac{99}{32}\frac{(\omega')^2\omega''}{\omega^6}
-\frac{13}{32}\frac{(\omega'')^2}{\omega^5}
-\frac{5}{8}\frac{\omega'\omega'''}{\omega^5}
+\frac{1}{16}\frac{\omega''''}{\omega^4}.
\end{align}

In \cite{Parker:1974qw} it was argued that, for a sufficiently slow and
smooth expansion, it is the choice $\b=0$ for the mode function 
\bea
\vphi(\tau,k)
=\frac{1}{\sqrt{2W}}
e^{-i\int^\tau W(\tau') d\tau'},
\label{wkbscalar}
\eea 
which
give rises to operators $a_\vk$ that corresponds to physical
particles. This choice of the vacuum 
\be
a_{\vk} \ket{0}_A =0.
\ee
is called the adiabatic vacuum.
Note that for a time independent metric, all higher order terms vanish
and $W = \o$. In this case the mode function $\vphi(\t,k)$ 
is the ordinary positive
frequency solution and the adiabatic vacuum reduces to the 
standard Minkowski vacuum.

In the adiabatic regularization, 
the renormalized energy momentum tensor is given by
\bea
\langle T_{\mu\nu} \rangle_{\rm ren}
= \langle T^{(m=0)}_{\mu\nu} \rangle - \lim_{m\to 0} {}_A
\langle 0| T_{\mu\nu} |0\rangle_A.
\label{regscalar}
\eea
As the adiabatic expansion
becomes more accurate for large $k$,
the second adiabatic subtraction term has the same UV divergent
structure as that of the first term and so 
$\langle T_{\mu\nu} \rangle_{\rm ren}$ is UV finite.
It is known that the adiabatic
regularization of the energy momentum tensor is equivalent to
renormalizing the gravitational coupling constants in the Einstein
equation \cite{Bunch:1980vc}. 
% we have to put a reference of Parker Fulling.
%yoji0930
As a matter of fact, the adiabatic regularization is a method for
renormalization rather than that for regularization of divergent
momentum integrals, since the adiabatic subtraction term  precisely
cancels mode by mode the contribution from large momenta to the first
term in \eqref{regscalar}, and the result $\langle T_{\mu\nu}
\rangle_{\rm ren}$ is thus completely finite.  
In order to remove all the divergences in the expectation value of the
energy momentum tensor, the adiabatic expansion should be performed up
to the fourth order, i.e. the same order as the mass dimension of
$T_{\m\n}$, the physical quantity being considered.

For our theory, 
substituting $\omega=k^2+m^2C$ to \eqref{adiabaticscalar}, 
the adiabatic expansion for $W$ up to  the fourth adiabatic order is
given by 
\begin{align}
W
&=
\omega
-\frac{m^2C''}{8\omega^3}+\frac{5m^4(C')^2}{32\omega^5}
+\frac{m^4 C''''}{32\omega^5}
-\frac{m^4}{128\omega^7}\Big(28C'''C'+19(C'')^2\Big)
\notag\\
&\quad
+\frac{221m^6C''(C')^2}{256\omega^9}
-\frac{1105m^8(C')^4}{2048\omega^{11}},
\end{align}
where we have absorbed back the formal expansion parameter $\e$ into
the time derivatives, effectively setting $\epsilon=1$.
The conformal anomaly in the classically conformally invariant theory
is determined by the massless limit of the adiabatic subtraction term 
\begin{align}
\langle T^\mu_{\ \mu} \rangle_{\rm ren}
&
= 
- \lim_{m\to 0} {}_A\langle 0| T^\mu_{\ \mu} |0\rangle_A
\notag\\
&
=
- \lim_{m\to 0} \frac{m^2}{4\pi^2C}
\int^{\infty}_{0} dk \frac{k^2}{W_k}
\notag\\
&
=
- \lim_{m\to 0} \frac{m^2}{4\pi^2C}
\int^{\infty}_{0} dk k^2
\left[
\frac{1}{\omega}
+\frac{m^2C''}{8\omega^5}-\frac{5m^4(C')^2}{32\omega^7}
-\frac{m^4 C''''}{32\omega^7}
\right.
\notag\\
&\qquad
\left.
-\frac{m^4}{128\omega^9}\Big(28C'''C'+21(C'')^2\Big)
-\frac{231m^6C''(C')^2}{256\omega^{11}}
+\frac{1155m^8(C')^4}{2048\omega^{13}}
\right]
\notag\\
&
=
\frac{1}{960\pi^2C^2}
\Big(
5(C')^4 - 11C(C')^2C'' + 3C^2(C'')^2
+ 4C^2C'C''' -C^2C''''
\Big)
\notag\\
&
=
\frac{1}{2880\pi^2}
\left[
\Big(R_{\mu\nu}R^{\mu\nu}-\frac13 R^2\Big)
+\Box R
\right],
\label{canomalyscalar}
\end{align}
where we have used \eqref{r2} in the last equality. In 
the above computation only the fourth adiabatic order terms survive to
contribute. 
%yoji0930
%c1 It is because that, 
The reason is that by introducing an UV momentum cutoff $k =
a(\tau)\Lambda$ where $\Lambda$ is the physical momentum cutoff, the
first and second adiabatic order terms give contributions which are
proportional to $m^4$ and $m^2$, respectively, and thus they vanish by
taking $m\to 0$ before taking $\Lambda\to \infty$. Note that the conformal
anomaly can be expressed in terms of the Ricci tensor and the scalar
curvature only since the Weyl tensor $C_{\mu\nu\rho\sigma}$ is
identically zero in a conformally flat spacetime
\cite{Parker:2009uva}.
%
%yoji0930
The conformal anomaly obtained here \eqref{canomalyscalar} agrees with
the result obtained by the other regularization methods
\cite{Dowker:1976zf,Duff:1977ay,Brown:1976wc,Bunch:1978yq}. 
We note that 
one has to be aware of the sign difference in front of the $\Box R$ term in
comparing the result here with those in
\cite{Dowker:1976zf,Bunch:1978yq}, which is simply due to the
convention of metric and the curvature tensors. For example, Dowker et
al. \cite{Dowker:1976zf} adopted 
$\eta_{\mu\nu}={\rm diag} (1,-1,-1,-1)$, ${}^{\rm D}R_{\mu\nu\sigma}^{\ \ \ \ \rho}
=\partial_\mu\Gamma^{\rho}_{\ \nu\sigma}-\cdots$ 
and ${}^{\rm D}R_{\nu\sigma}={}^{\rm D}R_{\rho\sigma\nu}^{\ \ \ \ \rho}$.
Therefore we have $R_{\mu\nu}={}^{\rm D}R_{\mu\nu}$ and the sign difference
comes from the metric sign convention as $\Box R = - \Box {}^{\rm D}
R$. The same remark applies to the results \eq{canomalyfermion} and
\eqref{canomaly2}.
%yoji1006 
%c2
% The another remark on the $\Box R$ term in the conformal anomaly 
% should be mentioned here is a fact that this term can be expressed 
% as the variation of $R^2$ term of an action \cite{Brown:1977pq,Duff:1977ay}: 
% \bea \sqrt{-g} \Box R =
% \frac{1}{6 }g^{\mu\nu}\frac{\delta}{\delta g^{\mu\nu}}
% \int d^4x \sqrt{-g} R^2,
% \label{boxr}
%\eea
%where the term on the right hand side is written explicitly as the
%form of a contribution  to the trace of the energy momentum tensor. 
% It is a contrast to the other terms
% appearing in the conformal anomaly that they can not be expressed as 
% the variation of local terms of an action. 
% Then, the coefficient 
% of $\Box R$ term can actually be set to any number by adding 
% the local finite counter-term ($R^2$ term) into the action. 
% Thus the ambiguity in the coefficient of $\Box R$ can also be found
% in the renormalization procedure. 
% For further discussions on the indeterminacy of $\Box R$ term, 
% see \cite{Vieira:2015oka}.
%

%%%%%%%%%%%%%%%%%%%%%%%%%%%%%%%%%%%%%%%%%%%%%%%%%%%%%%%%%%%%%%%%%%%%%%%%%%%%%%%%%%%%%%%%
\subsection{Dirac fermion}
%%%%%%%%%%%%%%%%%%%%%%%%%%%%%%%%%%%%%%%%%%%%%%%%%%%%%%%%%%%%%%%%%%%%%%%%%%%%%%%%%%%%%%%%

%yoji0930
Adiabatic expansion for a Dirac fermion has been performed recently 
in \cite{Landete:2013lpa,delRio:2014cha}. This is noticeably different
from the scalar field case which is based on the WKB type expansion.
As we are ultimately interested in the de-Sitter space superconformal
Yang-Mills theory \cite{Chu:2016uwi}, 
we will follow their convention and
consider the action for a Dirac fermion $\Psi$ in the form 
\be
S
= \int d^4 x
\sqrt{-g}
\Big(
-{\bar \Psi}\gamma^\mu(\partial_\mu+
\frac{1}{4}\omega_{\mu{\hat \rho}{\hat \sigma}}\gamma^{{\hat \rho}{\hat \sigma}})
\Psi
+m{\bar \Psi}\Psi
\Big),
\ee
where ${\bar \Psi}=i\Psi^\dagger \gamma^0$,
\bea
\gamma^{\hat 0}
=
\begin{pmatrix}
0 & 1 \\
-1 & 0
\end{pmatrix},
\qquad
\gamma^{\hat i}
=
\begin{pmatrix}
0 & \sigma^{\hat i} \\
\sigma^{\hat i} & 0
\end{pmatrix}.
\eea 
and hatted indices are those for the Minkowski space.
Note that the spin connection for a conformally flat spacetime 
is given by $\omega_{\mu{\hat \rho}{\hat \sigma}}
=-\frac{1}{2}D(\delta^{ 0}_{\ {\hat \rho}}\eta_{{\hat
    \sigma}\mu}-\delta^{ 0}_{\ {\hat \sigma}}\eta_{{\hat \rho}\mu})$,
where $D\equiv C'/C=2a'/a$.
It is convenient to introduce the rescaled field 
$\psi \equiv a^{-\frac{3}{2}}\Psi$, ${\bar \psi}=i\psi^\dagger
\gamma^{\hat 0}$,  the field equation for the rescaled field becomes
\bea
(\gamma^{\hat \mu}\partial_{\hat \mu}
-
m a)\psi =0,
\label{feqfermion0}
\eea 
%yoji0930
which is simply the free field equation for a Dirac fermion in
Minkowski space with a time dependent mass.

The Fourier expansion for $\psi$ is given by
\bea
\psi
=
\sum_{s=1,2}
\int\frac{d^3k}{(2\pi)^3}
\Big(
c_{\vec k}^su^s_{\vec k}(\tau)e^{i{\vec k}\cdot{\vec x}}
+d_{\vec k}^{s\dagger}v^s_{\vec k}(\tau)e^{-i{\vec k}\cdot{\vec x}}
\Big)
\eea
where $s$ is spin index and $v^s_{\vec k}$ is 
given by the charge conjugation of $u^s_{\vec k}$, $v^s_{\vec k}=u^{s
  C}_{\vec k}=\gamma^{\hat 0}\gamma^{\hat 1}\gamma^{\hat
  3}u^{s\ast}_{\vec k}$. The canonical anti-commutation relations are
\bea
\{\psi_\alpha(\tau,{\vec x}),\psi^{\dagger}_\beta(\tau,{\vec y})\}
=-\delta_{\alpha\beta}\delta^{(3)}({\vec x}-{\vec y}),
\label{comfermion}
\eea
where $\alpha$, $\beta$ are spinor indices.
%yoji0930
In addition to the field equation, the spinor mode functions are
subject to the orthogonality condition, 
\bea
\sum_{\alpha}u^{s\dagger}_{{\vec k} \alpha}(\tau)u^{s'}_{{\vec k} \alpha}(\tau)
=\delta^{ss'},
\label{ortho}
\eea
which guarantees the correct normalization of the scalar product of
$\Psi$ \cite{Greiner:1996zu}.
%
%c1 
Following \cite{Landete:2013lpa}, we write the spinor mode function
$u^s_{\vec k}$ as
\bea
u^s_{\vec k}
=
\begin{pmatrix}
h^I_k(\tau,\lambda_s)\xi^s
\\
h^{II}_k(\tau,\lambda_s)\xi^s
\end{pmatrix}
.
\label{modefermion}
\eea
Here $\xi^s$ is a two component spinor satisfying 
%yoji0930
\bea
\sum_{\alpha}\xi^{s\dagger}_{\alpha}\xi^{s'}_{\alpha}=\delta^{ss'},
\qquad
\sum_{s}\xi^{s}_{\alpha}\xi^{s\dagger}_{\beta}=\delta_{\alpha\beta},
\qquad
\frac{{\vec \sigma}\cdot{\vec k}}{k}\, \xi^s=\lambda_s \xi^s,
\eea
with $\lambda_s=\pm 1$ the helicity eigenvalues, and $h_k^{I,II}$ are 
scalar functions depending on $\lambda_s$. 
Substituting \eqref{modefermion} into 
%yoji0930
the field equation 
\eqref{feqfermion0}, we obtain 
\bea
(\partial_0+i\lambda_s k)h^{II}_k=m a h^I_k,
\qquad
(-\partial_0+i\lambda_s k)h^{I}_k=m a h^{II}_k,
\label{feqfermion1}
\eea
and it follows from the above equations that the second order
differential equations
\bea
\Big(\partial_0^2-\frac{D}{2}\partial_0
+k^2+m^2C+i\frac{D}{2}\lambda_s k\Big)h^{I}_k=0,
\notag
\\
\Big(\partial_0^2-\frac{D}{2}\partial_0 +k^2+m^2C-i\frac{D}{2}
\lambda_s k\Big)h^{II}_k=0.
\label{feqfermion2}
\eea
%Here explain how WKB does not work for fermion.
Eliminating the first derivative terms by redefining 
$h^{I,II}_k = a^{1/2}{\tilde h}^{I,II}_k$ yields
\bea
\Big(\partial_0^2+\Omega^2_{F}+i\frac{D}{2}\lambda_s k\Big){\tilde
  h}^{I}_k=0,
\qquad
\Big(\partial_0^2+\Omega^2_{F}-i\frac{D}{2}\lambda_s k\Big){\tilde
  h}^{II}_k=0,
\label{feqfermion3}
\eea
where 
\be 
\Omega^2_{F} = \omega^2 +\frac{D'}{4}-\frac{D^2}{16},\quad \mbox{and}
\quad \omega^2=k^2+m^2C.
\ee 

%c1 
It is important to note that the orthogonality condition 
\eq{ortho} implies the normalization condition for the scalar
functions 
\bea
|\htt^{I}_k(\tau,\lambda_s)|^2+|\htt^{II}_k(\tau,\lambda_s)|^2=1/a.
\label{nofermion}
\eea
%
%From \eqref{comfermion} and \eqref{nofermion} we have
%\bea
%\{c^s_{\vec k},c^{s'\dagger}_{{\vec k}'}\}
%=
%\{d^s_{\vec k},d^{s'\dagger}_{{\vec k}'}\}
%=
%-\delta^{ss'}(2\pi)^3\delta^{(3)}({\vec k}-{\vec k}').
%\eea
%yoji0930
%c1 Now let us consider the adiabatic expansion for the scalar functions
% $h^{I,II}_k$. 
It is obvious that a simple form of ansatz
\eqref{wkbscalar} like that for the scalar field could not solve the
normalization condition \eqref{nofermion}. As demonstrated in
\cite{Landete:2013lpa}, one needs to adopt an ansatz where the
amplitudes and the phases of the mode functions are
independent. Expanding adiabatically, the correct WKB type solution
for a Dirac fermion is of the form
\begin{align}
h^{I}_{k(n)}
&=
\sqrt{\frac{\omega-\lambda_sk}{2\omega}}(1+F_{(1)}+\cdots+F_{(n)})
\, e^{-i\int^\tau (\omega+\omega_{(1)}+\cdots+\omega_{(n)})d\tau'},
\notag\\
h^{II}_{k(n)}
&=
i\sqrt{\frac{\omega+\lambda_sk}{2\omega}}(1+G_{(1)}+\cdots+G_{(n)})
\, e^{-i\int^\tau (\omega+\omega_{(1)}+\cdots+\omega_{(n)})d\tau'}.
\end{align}
With these ansatz, one can obtain $\omega_{(n)}$, $F_{(n)}$ and $G_{(n)}$ by solving \eqref{feqfermion1} and \eqref{nofermion} iteratively. 
%yoji0930
Note that in the adiabatic expansion for a Dirac fermion, 
%c1 
terms of all
adiabatic order ($n=0,1,2,\cdots$) exist unlike the scalar field case 
where only terms of even adiabatic order are present.
%c1 the odd adiabatic order terms are absent.  
%
Here we will not repeat the same procedure for the 
adiabatic expansion and the adiabatic regularization of the energy
momentum tensor for a Dirac fermion field as it 
%c1 
has been carried out
in details in
\cite{Landete:2013lpa,delRio:2014cha}. The result for the conformal
anomaly agrees with the result obtained by the other regularization
methods \cite{Dowker:1976zf,Duff:1977ay}
\begin{align}
\langle T^\mu_{\ \mu} \rangle_{\rm ren}
=
\frac{1}{2880\pi^2}
\left[
11\Big(R_{\mu\nu}R^{\mu\nu}-\frac13 R^2\Big)
+6\Box R
\right].
\label{canomalyfermion}
\end{align}

%%%%%%%%%%%%%%%%%%%%%%%%%%%%%%%%%%%%%%%%%%%%%%%%%%%%%%%%%%%%%%%%%%%%%%%%%%%%%%%%%%%%%%%%
\section{Adiabatic expansion for $U(1)$ gauge field}
%%%%%%%%%%%%%%%%%%%%%%%%%%%%%%%%%%%%%%%%%%%%%%%%%%%%%%%%%%%%%%%%%%%%%%%%%%%%%%%%%%%%%%%%

As we mentioned above, 
the adiabatic expansion for gauge field has not been performed before
in the literature. As the theory has gauge symmetry, one needs to fix
a gauge in order to perform perturbative calculations.
%yoji0930
First thing to be clarified is what kind of gauge fixing term should
be used. Since the classical action of $U(1)$ gauge theory on a 
conformally flat spacetime in 4 dimensions possesses conformal 
invariance, one may
think that it is useful to adopt a gauge fixing term which preserves
the classical conformal invariance  
\bea
\cL_{\rm gf} = - \frac{ \sqrt{-g}}{2}  (\partial^\mu A_{\mu})^2.
\label{gfnonc}
\eea 
Using \eqref{gfnonc} the gauge fixed action with the ghost kinetic
term is conformally invariant and can be written precisely as the same
form as that in flat Minkowski space. In this case, the gauge field
and the ghost fields are simply described, 
%c1
respectively, 
by collections of 4 and 2
massless conformally coupled scalar modes. As a result,
the conformal anomaly in the adiabatic regularization amounts to
$(4-2)\times \langle T^\mu_{\ \mu} \rangle^{\rm scalar}_{\rm
  ren}$. 
%c1
This is wrong. 
%c1 The origin of having this undesired result for the conformal anomaly 
The reason why this gives the wrong result
is because the gauge fixing term \eqref{gfnonc} breaks the
general covariance and this leads to the breaking of the covariant
conservation of the energy momentum tensor. In this case it is thus
impossible to identify the pure conformal anomalous contribution to
the expectation value of the trace of energy momentum
tensor. Therefore, in order to evaluate the conformal anomaly
correctly, we have to use a gauge fixing term that respects the
general covariance even though by itself it breaks the classical conformal
invariance of the theory. 
Taking into account of the above 
%c1 argument, 
consideration, 
we will take the following 
covariant gauge fixing term with a parameter $\xi$, 
\bea
\cL_{\rm gf}=-\frac{\sqrt{-g}}{2\xi} (\nabla^{\mu}A_{\mu})^2.
\eea

%yoji0930
In order to perform the adiabatic expansion for the mode functions in
the $U(1)$ gauge theory, we introduce a mass $m$ for the
gauge field and a mass 
$m_\chi$ for  the (anti-)ghost fields $\chi$, ${\bar \chi}$,
respectively in such a way that the gauge-fixed massless $U(1)$ gauge
theory is recovered in the limit $m,m_{\chi}\to 0$ \cite{Dowker:1976zf}. 
The Lagrangian to be considered is thus
\begin{align}
\cL=\sqrt{-g}
\Big(
-\frac14 g^{\mu\rho}g^{\nu\sigma}F_{\mu\nu}F_{\rho\sigma}
-\frac{1}{2\xi} (\nabla^{\mu}A_{\mu})^2
-\frac12 m^2 g^{\mu\nu}A_{\mu}A_{\nu}
-i{\bar \chi}g^{\mu\nu}\nabla_\mu\nabla_\nu\chi
+i m_{\chi}^2 {\bar \chi}{\chi}
\Big),
\label{lag}
\end{align}
where
$F_{\mu\nu}=\nabla_{\mu}A_{\nu}-\nabla_{\nu}A_{\mu}=\partial_{\mu}
A_{\nu}-\partial_{\nu}A_{\mu}$.
The field equations derived from \eqref{lag} are
\bea
&& \hspace{-0.5cm}
\eta^{\rho\sigma}\partial_{\rho}\partial_\sigma A_{\mu}
+\Big(\frac{1}{\xi}-1\Big)\eta^{\rho\sigma}\partial_\mu\partial_\rho 
A_{\sigma}
-m^2CA_{\mu}
\notag\\
&&
+\frac{1}{\xi}
\Big[\delta^{\ 0}_{\mu}\Big(
-D\eta^{\rho\sigma}\partial_{\rho} A_{\sigma}
+D^2A_0 - D'A_0 \Big) 
- D\partial_\mu A_0
\Big]
=0,
\label{feqg}
\eea
and
\bea
\eta^{\mu\nu}\partial_\mu\partial_\sigma \chi - D\partial_0 \chi - 
m_{\chi}^2C \chi=0,\quad \text{same for }{\bar \chi}.
\label{feqgh}
\eea
Here $D=C'/C$ and $C=a^2$ as before.

Next we quantize the theory \eqref{lag} in the canonical formalism. 
First of all, the canonical conjugate momenta are defined by
\begin{align}
\pi_{A}^\mu 
&= \frac{\partial \cL}{\partial\partial_0 A_\mu}
= 
\eta^{\mu\nu}(\partial_0A_\nu-\partial_\nu A_0)
-\frac{1}{\xi}\eta^{\mu 0}(\eta^{\alpha\beta}\partial_\alpha A_\beta-DA_0),
\label{camg}
\\
%yoji0930
\pi_{\chi} 
&
=\frac{\partial \cL}{\partial\partial_0\chi} 
=-iC\partial_0{\bar \chi},
\qquad 
\pi_{\bar \chi}
=\frac{\partial \cL}{\partial\partial_0{\bar \chi}} 
=iC\partial_0{\chi}.
\label{camgh}
\end{align}
%yoji0930
In terms of \eqref{camg}, the temporal and spatial components of the
field equation for the gauge field are written as
\bea
&&
-\partial_i\pi_A^i +(\partial_0-D)\pi^0_A -m^2 C A_0=0,
\label{feqpa1}
\\
&&
%c1 
- \delta_{ik}\partial_0\pi_A^k
+\delta^{jk}\partial_j(\partial_kA_i-\partial_iA_k)
+\partial_i\pi^0_A -m^2 C A_i=0,
\label{feqpa2}
\eea
respectively.
%yoji0930
In order to decouple the field equations \eqref{feqpa1} and
\eqref{feqpa2}, we follow the strategy of \cite{Frob:2013qsa} and
separate the canonical variables into the transverse and the
longitudinal parts,
\bea
A_i = B_i+\partial_iA, \qquad
\pi_A^i = \delta^{ij} (w_j+\partial_j\pi_A),
\label{dec}
\eea
with $\partial^iB_i=\partial^iw_i=0$. 
Substituting the decompositions \eqref{dec} into the field 
equations \eqref{feqpa1} and \eqref{feqpa2} and using \eqref{camg}, we
arrive at three decoupled equations for $B_i$, $\pi_A^0$ and $\pi_A$, 
%yoji0930
\bea
&&
(\partial_0^2- \partial_j^2 +m^2C)B_i=0,
\label{feq1}
\\
&&
(\partial_0^2 - \partial_j^2 -D\partial_0+\xi m^2C-D')\pi^0_A=0,
\label{feq2}
\\
&&
(\partial_0^2 - \partial_j^2 -D\partial_0 + m^2C) \pi_A=0,
\label{feq3}
\eea
where $\partial_j^2 := \delta^{jk}\partial_j\partial_k$.
%c1  and find that
$w^i$  turns out to be  a dependent variable,
\bea
w_i=\partial_0B_i,
\label{btowi}
\eea
and 
%c1 
$A_0$ and $A$ can be obtained by using \eqref{feqpa1} and
\eqref{feqpa2} as
%yoji0930
\begin{align}
A_0 
= \frac{1}{m^2C}\Big((\partial_0-D)\pi^0_A- \partial_j^2\pi_A\Big),
\qquad
A 
= \frac{1}{m^2C}(\pi^0_A-\partial_0\pi_A).
\label{pitoa}
\end{align}
%

%yoji0930
The canonical (anti-)commutation relations are
\bea
&&[A_{\mu}(\tau,{\vec x}),\pi^{\nu}_A(\tau,{\vec x}')]
=i\delta^\nu_{\ \mu}\delta^{(3)}({\vec x}-{\vec x}'),
\label{commutatora}
\\
&&\{\chi(\tau,{\vec x}),\pi_{\chi}(\tau,{\vec x}')\}
=i\delta^{(3)}({\vec x}-{\vec x}'),
\qquad
\{{\bar \chi}(\tau,{\vec x}),\pi_{\bar \chi}(\tau,{\vec x}')\}
=i\delta^{(3)}({\vec x}-{\vec x}'),
\label{commutatorchi}
\eea
%c1
with the other (anti-)commutators vanish. 
The Fourier expansions of the dynamical variables 
$B_i$, $\pi^0_A$, $\pi_A$, $\chi$ and ${\bar \chi}$ are given by 
\begin{align}
B_i(\tau,{\vec x})
&=
\int\frac{d^3k}{(2\pi)^3}\sum_{{\rm p}=1,2}
\Big(\epsilon_i^{\rm p}
({\vec k})a_{\vec k}^{({\rm p})}f^{({\rm p})}(\tau,k)e^{i{\vec k}\cdot{\vec x}}
+\text{h.c}\Big),
\label{bi}
\\
\pi_A^0(\tau,{\vec x})
&=
\int\frac{d^3k}{(2\pi)^3}
\Big(a_{\vec k}^{(0)}f^{(0)}(\tau,k)e^{i{\vec k}\cdot{\vec x}}
+\text{h.c}\Big),
\label{pa0}
\\
\pi_A(\tau,{\vec x})
&=
\int\frac{d^3k}{(2\pi)^3}
\Big(a_{\vec k}^{(3)}f^{(3)}(\tau,k)e^{i{\vec k}\cdot{\vec x}}
+\text{h.c}\Big),
\label{pa}
\\
\chi(\tau,{\vec x})
&=
\int\frac{d^3k}{(2\pi)^3}
\Big(b_{\vec k}\chi(\tau,k)e^{i{\vec k}\cdot{\vec x}}
+b^{\dagger}_{\vec k}\chi^{\ast}(\tau,k)e^{-i{\vec k}\cdot{\vec x}}\Big),
\label{chi}
\\
{\bar \chi}(\tau,{\vec x})
&=
\int\frac{d^3k}{(2\pi)^3}
\Big({\bar b}_{\vec k}{\bar \chi}(\tau,k)e^{i{\vec k}\cdot{\vec x}}
+{\bar b}^{\dagger}_{\vec k}{\bar \chi}^{\ast}(\tau,k)e^{-i{\vec k}\cdot{\vec x}}\Big),
\label{bchi}
\end{align}
where $\epsilon_i^{\rm p}({\vec k})$ is the polarization tensor of the
transverse modes which satisfies 
\bea
\sum_i k^i\epsilon_i^{\rm p}({\vec k})=0,
\qquad
\sum_i \epsilon_i^{\rm p}({\vec k})\epsilon_i^{{\rm p}'}({\vec k})=\delta^{{\rm p}{\rm p}'},
 \qquad 
\sum_{{\rm p}=1,2}\epsilon_i^{\rm p}({\vec k})\epsilon_j^{\rm p}({\vec k})
=\delta_{ij}-\frac{k_ik_j}{k^2}.
\notag
\eea
According to \eqref{btowi} and \eqref{pitoa}, 
the corresponding Fourier expansions for $w_i$, $A_0$ and $A$ are
obtained as
\begin{align}
w_i(\tau,{\vec x})
&=
\int\frac{d^3k}{(2\pi)^3}\sum_{{\rm p}=1,2}
\Big(\epsilon_i^{\rm p}({\vec k})a_{\vec k}^{({\rm
    p})}\partial_0f^{({\rm p})}(\tau,k)e^{i{\vec k}\cdot{\vec x}}
+\text{h.c}\Big),
\label{wi}
\\
A_0(\tau,{\vec x})
&=
\frac{1}{m^2C}\int\frac{d^3k}{(2\pi)^3}
\Big(a_{\vec k}^{(0)}(\partial_0-D)f^{(0)}(\tau,k)e^{i{\vec k}\cdot{\vec x}}
+a_{\vec k}^{(3)}k^2f^{(3)}(\tau,k)e^{i{\vec k}\cdot{\vec x}}
+\text{h.c}\Big)
,
\label{a0}
\\
A(\tau,{\vec x})
&=
\frac{1}{m^2C}\int\frac{d^3k}{(2\pi)^3}
\Big(a_{\vec k}^{(0)}f^{(0)}(\tau,k)e^{i{\vec k}\cdot{\vec x}}
-a_{\vec k}^{(3)}\partial_0f^{(3)}(\tau,k)e^{i{\vec k}\cdot{\vec x}}
+\text{h.c}\Big).
\label{a}
\end{align}
Now we substitute \eqref{bi} -- \eqref{a} into 
\eqref{commutatora} and \eqref{commutatorchi}
to solve for 
the canonical (anti-)commutation relations for the creation
and annihilation operators and 
%c1 
the normalization  condition for the mode functions. 
%c1 It turns out that
We obtain 
\bea
[a^{({\mu})}_{\vec k},a^{(\nu)\dagger}_{{\vec k}'}]
=\eta^{{\mu}{\nu}}(2\pi)^3\delta^{(3)}({\vec k}-{\vec k}'),
\quad
\{b_{\vec k},{\bar b}^{\dagger}_{{\vec k}'}\}
=-\{{\bar b}_{\vec k},b^{\dagger}_{{\vec k}'}\}
=i(2\pi)^3\delta^{(3)}({\vec k}-{\vec k}'),
\label{commutator2}
\eea
where $\mu,\nu=0,1,2,3$, 
and the following normalization conditions for the mode functions
\bea
&&
f^{(1,2)}(\tau,k)\partial_0f^{(1,2)\ast}(\tau,k)-\partial_0f^{(1,2)}(\tau,k)f^{(1,2)\ast}(\tau,k)=i,
\notag\\
&&
f^{(0)}(\tau,k)\partial_0f^{(0)\ast}(\tau,k)-\partial_0f^{(0)}(\tau,k)f^{(0)\ast}(\tau,k)
=im^2C,
\notag\\
&&
f^{(3)}(\tau,k)\partial_0f^{(3)\ast}(\tau,k)-\partial_0f^{(3)}(\tau,k)f^{(3)\ast}(\tau,k)
=im^2Ck^{-2},
\notag\\
&&
\chi(\tau,k)\partial_0{\bar \chi}^{\ast}(\tau,k)-\partial_0{\bar \chi}(\tau,k)\chi^{\ast}(\tau,k)
=iC^{-1},
\notag\\
&&
{\bar \chi}(\tau,k)\partial_0{\chi}^{\ast}(\tau,k)-\partial_0{\chi}(\tau,k){\bar \chi}^{\ast}(\tau,k)
=iC^{-1}.
\label{normalization}
\eea 
%yoji0930
In terms of the mode functions, the field equations \eqref{feqgh},
\eqref{feq1} -- \eqref{feq3} read
\begin{align}
&
(\partial_0^2+\omega^2)f^{(1,2)}(\tau,k)=0,
\label{feq4}
\\
&
(\partial_0^2-D\partial_0+\omega^2_0-D')f^{(0)}(\tau,k)=0,
\label{feq5}
\\
&
(\partial_0^2-D\partial_0+\omega^2)f^{(3)}(\tau,k)=0,
\label{feq6}
\\
&
(\partial_0^2-D\partial_0+\omega^2_\chi-D')\chi(\tau,k)=0,
\label{feq7}
\end{align}
where 
\be
\omega^2 :=k^2+m^2C, \quad \omega^2_0 :=k^2+\xi m^2C, \quad 
\omega^2_\chi :=k^2+ m_\chi^2C.
\ee

To perform the adiabatic expansion,
%c1  to solve \eqref{feq4} -- \eqref{feq7}. 
we  notice that 
the differential equations for $f^{(0)}(\tau,k)$, $f^{(3)}(\tau,k)$
and $\chi(\tau,k)$ include first time derivative terms that can be
eliminated by rescaling the mode functions by appropriate time
dependent functions. 
%c1
Defining
\begin{align}
f^{(0)}(\tau,k)
&=
(m^2C)^{\frac{1}{2}}Y_0(\tau,k),
\label{mode0}
\\
f^{(3)}(\tau,k)
&=
\left(\frac{m^2C}{k^2}\right)^{\frac{1}{2}}Y_L(\tau,k),
\label{mode3}
\\
\chi(\tau,k)
&=
C^{-\frac{1}{2}}Y_\chi(\tau,k)
,
\qquad \text{same for }  {\bar \chi}(\tau,k).
\label{modeg}
\end{align}
then the differential equations \eq{feq4}--\eq{feq7}  
simplify to the form
of a harmonic oscillator with a time dependent frequency,
\be \label{meq}
(\partial_0^2+\Omega^2_a)Y_a(\tau,k)=0,
\qquad (a=0,L,T,\chi),
\ee
where we have defined $Y_T(\tau,k) := f^{(1,2)}(\tau,k)$ , and
\be
\Omega_a^2 :=\omega_a^2 + \alpha_a,
\ee
with 
\bea
  \omega_a = \left\{ \begin{array}{ll}
    \omega_0 & (a=0) \\
   \omega & (a=L,T) \\
   \omega_\chi & (a=\chi)
  \end{array} \right.,
  \qquad
  \alpha_a = \left\{ \begin{array}{ll}
    -\frac{1}{6}CR & (a=0,\chi) \\
   \frac{1}{6}CR-\frac{1}{2}D^2 & (a=L) \\
   0 & (a=T) 
  \end{array} \right.,
\eea
and $R=C^{-1}(3D'+\frac{3}{2}D^2)$ being the scalar curvature. 
Note that the mode functions of the temporal component of the 
conjugate momentum $\pi^0_A$ and of the ghost field $\chi$  
satisfy the same
differential equation as that of a minimally coupled scalar field.
%c1
Note also that 
%for $\xi=1$ or $m=0$, 
we have 
\bea
Y_0(\tau,k)=Y_\chi(\tau,k), \qquad (m=m_\chi=0).
\label{tgeq}
\eea

At this point it is pleasing to note that the same rescaling also bring the
normalization conditions \eq{normalization} to the same
standard form \eq{normalizationscalar} as that of scalar field 
\be
Y_a\partial_0Y_a^{\ast}-\partial_0Y_aY_a^{\ast}=i \quad \mbox{(no sum
  over $a$)}.
\ee 
Therefore one can proceed to quantize the theory adiabatically
with the choice of the mode functions
\begin{align}
Y_a(\tau,k)
&=\frac{1}{\sqrt{2W_a(\tau)}}e^{-i\int^\tau W_a(\tau')d\tau'},
\label{wkby}
\end{align}
where
\bea
W^2_a
=\Omega^2_a
-\left(\frac{W''_a}{2 W_a}-\frac{3(W'_a)^2}
{4 W_a^2}\right);
\label{wkbequation}
\eea
and with the adiabatic vacuum $|0\rangle_A$  defined by
\bea
a^{({\mu})}_{\vec k}|0\rangle_A=b_{\vec k}|0\rangle_A={\bar b}_{\vec
  k}|0\rangle_A=0. 
\eea 

The adiabatic expansions are obtained by solving \eqref{wkbequation}
iteratively with the zeroth adiabatic order solutions
\bea
W_{a(0)}=\omega_a.
\label{wkbzeroth}
\eea
Then one obtains the following results up to the fourth adiabatic order,
\begin{align}
W_a
&=
\omega_a
-\frac{m_a^2C''}{8\omega_a^3}+\frac{5m^4(C')^2}{32\omega_a^5}
+\frac{\alpha_a}{2\omega_a}
+\frac{m_a^4 C''''}{32\omega_a^5}
-\frac{m_a^4}{128\omega_a^7}\Big(28C'''C'+19(C'')^2\Big) \label{wkbgauge1}
\\
&
+\frac{221m_a^6C''(C')^2}{256\omega_a^9}
-\frac{1105m_a^8(C')^4}{2048\omega_a^{11}}
-\frac{\alpha_a''}{8\omega_a^3}
+\frac{m_a^2}{16\omega_a^5}(5C'\alpha_a'+3C''\alpha_a)
-\frac{25m_a^4(C')^2\alpha_a}{64\omega_a^7} \nn
\end{align}
and 
\begin{align}
\frac{1}{W_a}
&=
\frac{1}{\omega_a}
+\frac{m_a^2C''}{8\omega_a^5}-\frac{5m_a^4(C')^2}{32\omega_a^7}
-\frac{\alpha_a}{2\omega_a^3}
-\frac{m_a^4 C''''}{32\omega_a^7}
-\frac{m_a^4}{128\omega_a^9}\Big(28C'''C'+21(C'')^2\Big) \label{wkbgauge2}\\
&
-\frac{231m_a^6C''(C')^2}{256\omega_a^{11}}
+\frac{1155m_a^8(C')^4}{2048\omega_a^{13}}
+\frac{\alpha_a''+3\alpha_a^2}{8\omega_a^5}
-\frac{5m_a^2}{16\omega_a^7}(C'\alpha_a'+C''\alpha_a)
+\frac{35m_a^4(C')^2\alpha_a}{64\omega_a^9}. \nn
\end{align}
Here we have introduced $m_a$ by $\o_a^2 = k^2 + m_a^2 C$. Explicitly,
it is 
\bea
m_a^2 = \left\{ \begin{array}{ll}
    \xi m^2 & (a=0) \\
   m^2 & (a=L,T) \\
   m_\chi^2 & (a=\chi)
  \end{array} \right. .
\eea 
The expressions \eqref{wkbgauge1} and \eqref{wkbgauge2}
have been expressed in ascending (even) powers of time derivatives.
%c1
The results obtained here for the 
adiabatic expansion of the $U(1)$ gauge field is new.
%

%%%%%%%%%%%%%%%%%%%%%%%%%%%%%%%%%%%%%%%%%%%%%%%%%%%%%%%%%%%%%%%%%%%%%%%%%%%%%%%%%%%%%%%%
\section{Adiabatic regularization of the energy momentum tensor and the conformal anomaly}
%%%%%%%%%%%%%%%%%%%%%%%%%%%%%%%%%%%%%%%%%%%%%%%%%%%%%%%%%%%%%%%%%%%%%%%%%%%%%%%%%%%%%%%%

Next let us turn to consider the adiabatic
regularization of the energy momentum tensor for the $U(1)$ gauge
theory \eqref{lag}. We will  focus on the conformal anomaly in this
paper.  The energy
momentum tensor obtained from \eqref{lag} is
\begin{align}
T_{\mu\nu}
&=
\frac{-2}{\sqrt{-g}}\frac{\delta S}{\delta g^{\mu\nu}}
\notag\\
&=
-\frac{1}{4}g_{\mu\nu}g^{\alpha\rho}g^{\beta\sigma}F_{\alpha\rho}F_{\beta\sigma}
+g^{\alpha\beta}F_{\alpha\mu}F_{\beta\nu}
-\frac{1}{2\xi} g_{\mu\nu} (g^{\alpha\beta}\nabla_\alpha A_\beta)^2
\notag\\
&\quad
+\frac{1}{\xi}(\nabla_{\mu}A_{\nu}+\nabla_{\nu}A_{\mu})
(g^{\alpha\beta}\nabla_\alpha A_\beta)
-\frac{1}{\xi}\Big[\nabla_{\mu}(A_\nu g^{\alpha\beta}\nabla_{\alpha}A_\beta)
+\nabla_{\nu}(A_\mu g^{\alpha\beta}\nabla_{\alpha}A_\beta)\Big]
\notag\\
&\quad
+\frac{1}{\xi}g_{\mu\nu}g^{\rho\sigma}
\nabla_{\rho}(A_\sigma g^{\alpha\beta}\nabla_{\alpha}A_\beta)
-\frac12g_{\mu\nu}m^2g^{\alpha\beta}A_{\alpha}A_{\beta}
+m^2A_\mu A_\nu 
\notag\\
&\quad
+ig_{\mu\nu}g^{\rho\sigma}\nabla_\rho{\bar \chi}\nabla_\sigma\chi
-i(\nabla_\mu{\bar \chi}\nabla_\nu\chi+\nabla_\nu{\bar \chi}\nabla_\mu\chi)
+ig_{\mu\nu}m_\chi^2{\bar \chi}\chi.
\label{stensor}
\end{align}
In the adiabatic regularization scheme, 
the renormalized energy momentum tensor is given by
\bea
\langle  T_{\mu\nu}(x) \rangle_{\rm ren}
= 
\langle T^{(m=0)}_{\mu\nu}(x) \rangle
-\lim_{m,m_\chi\to 0} 
{}_A\langle 0| T_{\mu\nu}(x)|0\rangle_{\rm A}.
\label{renstensor}
\eea
where the first term on the right hand side is evaluated in
%$|0\rangle$ is 
%given by the conformal 
the vacuum defined by the mode functions
in the massless theory.
%yoji0930
As we have explained before, in order to remove all the divergences
and obtain the finite result for the renormalized energy momentum
tensor, one should expand the subtraction term ${}_A\langle 0|
T_{\mu\nu}|0\rangle_{\rm A}$ up to the fourth adiabatic order. 

%
%${}_A\langle 0| T_{00}|0\rangle_{\rm A}$ at the fourth adiabatic order includes linear terms in $W_a$ which contains the term proportional to $\alpha''_a/\omega_a^3$ giving rise to logarithmic UV divergence. 
%

Now we evaluate the conformal anomaly.
Taking the trace of \eqref{stensor}, we obtain
\bea
T^\mu_{\ \mu}
= 
\frac{2}{\xi}g^{\mu\nu}\nabla_{\mu}(A_\nu g^{\alpha\beta}\nabla_{\alpha}A_\beta)
-m^2g^{\mu\nu}A_\mu A_\nu
+2ig^{\mu\nu}\partial_\mu{\bar \chi}\partial_\nu\chi
+4i m_\chi^2{\bar \chi}\chi.
\label{ctrace}
\eea
%c1
One may worry that \eqref{ctrace} does not vanish even in the
massless limit since the covariant gauge fixing term and the ghost
kinetic term breaks the conformal symmetry individually. However it is
easy to check that they indeed cancel each other and give zero
contribution  to the trace of the
energy momentum tensor when we take the expectation value with respect
to the vacuum defined in the massless theory where \eqref{tgeq}
holds:
\be \label{exactcancel}
\langle   T^{\mu}_{\ \mu}{}^{(m=0)}  \rangle =0.
\ee
As a result, the conformal anomaly is determined entirely 
by the adiabatic subtraction term
\bea
\langle   T^{\mu}_{\ \mu}  \rangle_{\rm ren}
= 
-\lim_{m,m_\chi\to 0} 
{}_A\langle 0| T^{\mu}_{\ \mu}|0\rangle_{\rm A}.
\label{canomaly}
\eea
%

%yoji0930
Let us start with the contribution from the mass term of the gauge
field in \eqref{ctrace}.
The corresponding adiabatic subtraction term which contributes to the
conformal anomaly is given by
\begin{align}
\langle  T^{\mu}_{\ \mu} \rangle_{\rm ren}^{\text{mass}} 
&=
- \lim_{m\to 0} 
{}_A\langle 0| (-m^2g^{\mu\nu}A_{\mu}A_\nu) |0\rangle_{\rm A}
\notag\\
&=
\lim_{m\to 0} \frac{m^2}{C}
\Big[-{}_A\langle 0| A^2_0|0\rangle_{\rm A}
+{}_A\langle 0|\delta^{ij} A_iA_j|0\rangle_{\rm A} \Big]
\notag\\
&=
\lim_{m\to 0} \frac{1}{C^{2}}
\int\frac{d^3k}{(2\pi)^3}
\left(
\Big|\Big(\partial_0-\frac{D}{2}\Big)Y_0(\tau,k)\Big|^2
-k^2 |Y_0(\tau,k)|^2
\right.
\notag\\
&
\quad
\left.
+
\Big|\Big(\partial_0+\frac{D}{2}\Big)Y_L(\tau,k)\Big|^2
-k^2|Y_L(\tau,k)|^2
%c1
+2m^2C|Y_T(\tau,k)|^2
\right).
\label{massterm}
\end{align}
Expanding this expression up to the  fourth
adiabatic order, we find that \eqref{massterm} is UV finite and 
gives a finite contribution to the
conformal anomaly. However, we found that the contribution from the
mass term alone cannot be expressed in terms of $R_{\mu\nu}^2$, $R^2$
and $\Box R$ only. The contribution from the other terms is thus
important to obtain the correct result.
%
%c1 
Next we evaluate the contribution from the term proportional to
$\xi^{-1}$ in \eqref{ctrace}, 
\begin{align}
\langle   T^{\mu}_{\ \mu}  \rangle_{\rm ren}^{\text{$\xi$}} 
&=
- \lim_{m\to 0} 
{}_A\langle 0| \frac{2}{\xi} g^{\mu\nu}\nabla_{\mu}(A_\nu
g^{\alpha\beta}\nabla_{\alpha}A_\beta) |0\rangle_{\rm A}
\notag\\
&=
-\lim_{m\to 0} \frac{2}{ C^2}
\Big[
{}_A\langle 0| \xi(\pi^0_A)^2|0\rangle_{\rm A}
-{}_A\langle 0| A_0(\partial_0-D)\pi^0_A|0\rangle_{\rm A}
+{}_A\langle 0| \delta^{ij}\partial_iA\partial_j\pi^0_A|0\rangle_{\rm A}
\Big]
\notag\\
&=
\lim_{m\to 0} \frac{2}{C^{2}}
\int\frac{d^3k}{(2\pi)^3}
\left(
-
\Big|\Big(\partial_0-\frac{D}{2}\Big)Y_0(\tau,k)\Big|^2
+
\omega_0^2|Y_0(\tau,k)|^2
\right).
\label{xidependent}
\end{align}
The fourth adiabatic order contribution from \eqref{xidependent} is
found to be UV divergent. 
%c1 Note that 
This UV divergence is canceled by
the ghost contribution as we will see below.
Finally the ghost contribution is obtained as
\begin{align}
\langle  T^{\mu}_{\ \mu}  \rangle_{\rm ren}^{\text{ghost}} 
&=
- 2i\lim_{m_\chi \to 0} 
{}_A\langle 0| (g^{\mu\nu}\partial_\mu{\bar \chi}\partial_\nu\chi
+2 m_\chi^2{\bar \chi}\chi) |0\rangle_{\rm A}
\notag\\
&=
\lim_{m_\chi \to 0} \frac{2}{C^{2}}
\int\frac{d^3k}{(2\pi)^3}
\left(
\Big|\Big(\partial_0-\frac{D}{2}\Big)Y_\chi(\tau,k)\Big|^2
-(k^2+2m_\chi^2C) |Y_\chi (\tau,k)|^2
\right).
\label{ghostterm}
\end{align}
For large $k$, taking into account of \eqref{tgeq}, we observe that
the two contributions from \eqref{xidependent} and \eqref{ghostterm}
have the same form but opposite sign, and so the 
%c1 
respective UV divergences cancel
each other to give a finite result in the conformal anomaly.
We remark that the expressions for the momentum integrations in \eqref{massterm} --
\eqref{ghostterm} are indeed valid for a general vacuum state until we
substitute the adiabatic expansions. As a result, the contributions
\eq{xidependent} and \eq{ghostterm} cancel exactly each other in the
massless theory and we obtain \eq{exactcancel}.

%

%c1 please check the subscript
Putting \eqref{massterm}, \eqref{xidependent} and \eqref{ghostterm}
together, 
the conformal anomaly for the $U(1)$ gauge theory in the adiabatic
regularization is given by
\begin{align}
\langle  T^{\mu}_{\ \mu}  \rangle_{\rm ren} 
&=
\lim_{m,m_\chi \to 0}
\frac{1}{4\pi^2C^2}\int^{\infty}_0 dk k^2 
\left[
\frac{2}{W_{0}}
\left(\omega_0^2 -\Big(\frac{W'_{0}}{2W_{0}}+\frac{D}{2}\Big)^2-W^2_{0}\right)
\right.
\notag\\
&\hspace{4cm}
-\frac{1}{W_{0}}
\left(k^2-\Big(\frac{W'_{0}}{2W_{0}}+\frac{D}{2}\Big)^2-W^2_{0}\right)
\notag\\
&\hspace{4cm}
-\frac{1}{W_{L}}
\left(k^2 -\Big(\frac{W'_{L}}{2W_{L}}-\frac{D}{2}\Big)^2-W^2_{L}\right)
+2\frac{m^2C}{W_T}
\notag\\
&\hspace{4cm}
\left.
-\frac{2}{{W}_{\chi}}
\left(k^2+2m_\chi^2C -\Big(\frac{{W}'_{\chi}}{2{W}_{\chi}}+\frac{D}{2}\Big)^2-{W}^2_{\chi}\right)
\right]_{(4)}
\notag\\
&=
\frac{1}{2880\pi^2}
\left[
-150\frac{(C')^4}{C^6}+474\frac{(C')^2C''}{C^5}-162\frac{(C'')^2}{C^4}
-216\frac{C'''C'}{C^4}+54\frac{C''''}{C^3}
\right.
\notag\\
&
\left.
\quad
-\log\xi
\left(\frac{405}{2}\frac{(C')^4}{C^6}-\frac{945}{2}\frac{(C')^2C''}{C^5}+135\frac{(C'')^2}{C^4}
+180\frac{C'''C'}{C^4}-45\frac{C''''}{C^3}
\right)
\right]
\notag\\
&=
\frac{1}{2880\pi^2}
\left[
62\Big(R_{\mu\nu}R^{\mu\nu}-\frac13R^2\Big)
-(18+15\log\xi)\Box R
\right],
\label{canomaly2}
\end{align}
where the subscript
$(4)$ denotes the term up to the fourth adiabatic order. In obtaining
this result,
we have used \eqref{wkbgauge1} and \eqref{wkbgauge2} in the 
second equality and \eqref{r2} in the third equality.
Note that the $\xi$ dependence came entirely from $W_0$ and $\o_0$ as
the other quantities $W_L, W_T, W_\chi$ are independent of $\xi$. 
%c2
The regularization independent term of our result \eqref{canomaly2} 
agrees precisely with that obtained (first term of \eq{other-v}) %yoji1010
using other regularization schemes. As for the regularization
dependent $\Box R$ term, a priori there is no need for our result to
agree with any of the previously obtained results. However to our
surprise, our value of $d$ agrees 
with the results of 
\cite{Dowker:1976zf} for $\xi=1$ 
obtained using zeta function regularization, 
and \cite{Endo:1984sz} for a general 
%c2 
gauge fixing parameter $\xi$
obtained using the DeWitt-Schwinger expansion.

%%%%%%%%%%%%%%%%%%%%%%%%%%%%%%%%%%%%%%%%%%%%%%%%%%%%%%%%%%%%%%%%%%%%%%%%%%%%%%%%%%%%%%%%
\section{Summary}
%%%%%%%%%%%%%%%%%%%%%%%%%%%%%%%%%%%%%%%%%%%%%%%%%%%%%%%%%%%%%%%%%%%%%%%%%%%%%%%%%%%%%%%%

In this article, we have investigated 
%c2
and constructed 
the adiabatic expansion and
regularization for a $U(1)$ gauge field in a 
conformally flat spacetime. This has never been considered 
%c1 carefully
before and our results are new. We argued the
necessity of the use of covariant gauge fixing term for the sake of
covariant conservation of the energy momentum tensor. 
Like in the scalar field case, the adiabatic expansion of the gauge
field mode functions are carried out by the WKB type solutions which
preserve the Wronskian type normalization conditions. 
%c2 We note 
It is clear that the adiabatic expansion and the computation of conformal
anomaly for a $U(1)$ gauge field performed here can be easily extended
to that for Yang-Mills gauge fields.

Based on the adiabatic expansion, we evaluated the conformal anomaly
for the $U(1)$ gauge field 
%c2 and found that, 
in a conformally flat spacetime;
%c2
and found that the result exactly agrees with that obtained from
$\zeta$ function regularization \cite{Dowker:1976zf,Endo:1984sz} in
the Dewitt-Schwinger (or local momentum expansion \cite{Bunch:1979uk})
formalism \cite{Birrell:1982ix}
%c1 , which can be applied to a general curved background 
and from the Hadamard renormalization
\cite{Belokogne:2015etf}.  
We have observed the same gauge dependence  in the coefficient of the $\Box
R$ 
%c2
term  of the conformal anomaly as eq. (5.1) of \cite{Endo:1984sz}.
However the result is different from that obtained 
using the dimensional regularization with $\xi=1$
\cite{Brown:1977pq,Duff:1977ay}.
%c2
Our result clearly confirms the regularization dependency of the $\Box R$
term of the conformal anomaly.

While we have focused on the conformal anomaly in this article,
evaluation of the renormalized energy momentum tensor (and more
general correlation functions) in a specific conformally flat 
spacetime, e.g. in de-Sitter space or in inflationary universe is an
important application of our adiabatic
regularization procedure. 
Since the adiabatic regularization allow one
to compute the particle number density, one can also discuss gauge
field particle production in an expanding universe.
Another important application is the
study of the renormalizability of the $\cN=4$ superconformal
Yang-Mills theory on de-Sitter space \cite{cy}.

%The conformal anomaly \eqref{canomaly2} does not make sense if we simply take $\xi \to \infty$ or $\xi\to 0$ in the above result. Ignoring the ghosts and taking the limit $\xi=\infty$ in \eqref{\lag} corresponds to the Proca theory. Actually, we have found that if we start with the Proca Lagrangian which is given by neglecting the ghost terms and taking $\xi\to \infty$ in \eqref{lag}, the adiabatic subtraction term for the quantum trace of the energy momentum tensor at the fourth adiabatic order has a logarithmic divergent contribution. This appearance of the divergence in the fourth adiabatic order subtraction term indicates that the expectation value of the energy momentum tensor of the Proca theory has a divergence that is absent in the $U(1)$ gauge theory.

\vskip7mm
%%%%%%%%%%%%%%%%%%%%%%%%%%%%%%%%%%%%%%%%%%%%%%%%%%%%%%%%%%%%%%%%%%%%%%%%%
\section*{Acknowledgements}
%%%%%%%%%%%%%%%%%%%%%%%%%%%%%%%%%%%%%%%%%%%%%%%%%%%%%%%%%%%%%%%%%%%%%%%%%

We would like to thank Bei-Lok Hu, Satoshi Iso and Yoshihisa Kitazawa 
for helpful discussions and comments. 
This work is
supported in part by the National Center of Theoretical Science
(NCTS) and the grant
104-2112-M-007-001 -MY3 of the Ministry of Science and
Technology of Taiwan.

%%%%%%%%%%%%%%%%%%%%%%%%%%%%%%%%%%%%%%%%%%%%%%%%%%%

\appendix

\section{Some geometrical tensors in conformally flat spacetime}
%%%%%%%%%%%%%%%%%%%%%%%%%%%%%%%%%%%%%%%%%%%%%%%%%%%%%%%%%%%%%%%%%%%%%%%%%%%%%%%%%%%%%%%%

%c1 The metric is given by \eqref{metric}. 
For a conformally flat spacetime \eq{metric}  in 4 dimensions, 
the Ricci tensor and Ricci scalar 
%in 4 dimensions 
are
\begin{align}
R_{\mu\nu}
&=
\frac32\delta^0_{\ \mu}\delta^0_{\ \nu}\left(\frac{C'}{C}\right)^2
+\frac{1}{2}(-2\delta^0_{\ \mu}\delta^0_{\ \nu}+\eta_{\mu\nu})\frac{C''}{C},
%\frac{D'}{2}((2-d)\delta^0_{\ \mu}\delta^0_{\ \nu}+\eta_{\mu\nu})
%+\frac{d-2}{4}D^2(\delta^0_{\ \mu}\delta^0_{\ \nu}+\eta_{\mu\nu}),
\notag\\
R
&=
C^{-1}\Big[-\frac32\left(\frac{C'}{C}\right)^2+3\frac{C''}{C}\Big].
%C^{-1}\Big[(d-1)D'+\frac14 (d-2)(d-1)D^2\Big].
\label{r1}
\end{align}
Quantities which appear at the fourth adiabatic order in a conformally
flat spacetime are 
\begin{align}
R_{\mu\nu}R^{\mu\nu}
&=
\frac{9}{4}\frac{(C')^4}{C^6}-\frac{9}{2}\frac{(C')^2C''}{C^5}+3\frac{(C'')^2}{C^4},
%C^{-2}\Big[\frac{d(d-1)}{4}(D')^2+\frac{(d-2)(d-1)}{4}D'D^2
%+\frac{(d-2)^2(d-1)}{16}D^4\Big],
\notag\\
R^2
&=
\frac{9}{4}\frac{(C')^4}{C^6}-9\frac{(C')^2C''}{C^5}+9\frac{(C'')^2}{C^4},
%C^{-2}\Big[(d-1)^2(D')^2+\frac{(d-2)(d-1)^2}{2}D'D^2
%+\frac{(d-2)^2(d-1)^2}{16}D^4\Big],
\notag\\
\Box R
&=
\frac{27}{2}\frac{(C')^4}{C^6}-\frac{63}{2}\frac{(C')^2C''}{C^5}+9\frac{(C'')^2}{C^4}
+12\frac{C'''C'}{C^4}-3\frac{C''''}{C^3}.
%\frac{1-d}{C^2}\Big[D'''+\frac14(d^2-11d+22) D'D^2
%+(d-4)\Big(D''D+\frac{(D')^2}{2}
%-\frac{d-2}{8}D^4\Big)\Big].
%\notag\\
\label{r2}
\end{align}

%%%%%%%%%%%%%%%%%%%%%%%%%%%%%%%%%%%%%%%%%%%%%%%%%%%%%%%%%%%%%%%%%%%%%%%%%%%%%%%%%%%%%%%%                            

%%%%%%%%%%%%%%%%%%%%%%%%%%%%%%%%%%%%%%%%%%%%%%%%%%%%%%%%%%%%%%%%%%%%%%%%%%%%%%%%%%%%%%%%


\begin{thebibliography}{100}
%%%%%%%%%%%%%%%%%%%%%%%%%%%%%%%%%%%%%%%%%%%%%%%%%%%%%%%%%%%%%%%%%%%%%%%%%%%%%%%%%%%%%%%%

\bibitem{Birrell:1982ix}
  N.~D.~Birrell and P.~C.~W.~Davies,
  ``Quantum Fields in Curved Space,''
  Cambridge University Press (1984). 
  %doi:10.1017/CBO9780511622632
  %%CITATION = doi:10.1017/CBO9780511622632;%%
  %1283 citations counted in INSPIRE as of 03 Oct 2016

\bibitem{Calzetta:2008iqa}
  E.~A.~Calzetta and B.~L.~B.~Hu,
  ``Nonequilibrium Quantum Field Theory,''
  Cambridge University Press (2008). 
  %%CITATION = INSPIRE-1384872;%%
  %2 citations counted in INSPIRE as of 03 Oct 2016


%\cite{Parker:2009uva}
\bibitem{Parker:2009uva} 
  L.~E.~Parker and D.~Toms, 
  ``Quantum Field Theory in Curved Spacetime : Quantized Field and Gravity,''
  Cambridge University Press (2009).
  %%CITATION = INSPIRE-1204522;%%
  %5 citations counted in INSPIRE as of 26 Sep 2016

%c2 
\bibitem{Chou:1984es} 
  K.~C.~Chou, Z.~B.~Su, B.~L.~Hao and L.~Yu,
  ``Equilibrium and Nonequilibrium Formalisms Made Unified,''
  Phys.\ Rept.\  {\bf 118}, 1 (1985).
  %%CITATION = PRPLC,118,1;%%
  %597 citations counted in INSPIRE as of 15 Nov 2014

\bibitem{Calzetta:1986ey} 
  E.~Calzetta and B.~L.~Hu,
  ``Closed Time Path Functional Formalism in Curved Space-Time:
  Application to Cosmological Back Reaction Problems,''
  Phys.\ Rev.\ D {\bf 35}, 495 (1987).
  %%CITATION = PHRVA,D35,495;%%
  %343 citations counted in INSPIRE as of 15 Nov 2014

\bibitem{Jordan:1986ug} 
  R.~D.~Jordan,
  ``Effective Field Equations for Expectation Values,''
  Phys.\ Rev.\ D {\bf 33}, 444 (1986).
  %%CITATION = PHRVA,D33,444;%%
  %295 citations counted in INSPIRE as of 15 Nov 2014
  
\bibitem{Weinberg:2005vy} 
  S.~Weinberg,
  ``Quantum contributions to cosmological correlations,''
  Phys.\ Rev.\ D {\bf 72}, 043514 (2005)
  [hep-th/0506236].
  %%CITATION = HEP-TH/0506236;%%
  %373 citations counted in INSPIRE as of 11 Oct 2014


\bibitem{dewitt}
B. S. DeWitt, 
{\it The Dynamical Theory of Groups and Fields}, Gordon and Breach,
New York, 1965.


%\cite{Dowker:1976zf}
\bibitem{Dowker:1976zf} 
  J.~S.~Dowker and R.~Critchley,
  ``The Stress Tensor Conformal Anomaly for Scalar and Spinor Fields,''
  Phys.\ Rev.\ D {\bf 16}, 3390 (1977).
  %doi:10.1103/PhysRevD.16.3390
  %%CITATION = doi:10.1103/PhysRevD.16.3390;%%
  %128 citations counted in INSPIRE as of 26 Sep 2016

%\cite{Hawking:1976ja}
\bibitem{Hawking:1976ja}
  S.~W.~Hawking,
  ``Zeta Function Regularization of Path Integrals in Curved Space-Time,''
  Commun.\ Math.\ Phys.\  {\bf 55} (1977) 133.
  doi:10.1007/BF01626516
  %%CITATION = doi:10.1007/BF01626516;%%
  %1023 citations counted in INSPIRE as of 01 Oct 2016

%\cite{Parker:1974qw}
\bibitem{Parker:1974qw} 
  L.~Parker and S.~A.~Fulling,
  ``Adiabatic regularization of the energy momentum tensor of a quantized field in homogeneous spaces,''
  Phys.\ Rev.\ D {\bf 9}, 341 (1974).
  %doi:10.1103/PhysRevD.9.341
  %%CITATION = doi:10.1103/PhysRevD.9.341;%%
  %266 citations counted in INSPIRE as of 26 Sep 2016

%\cite{Fulling:1974pu}
\bibitem{Fulling:1974pu} 
  S.~A.~Fulling, L.~Parker and B.~L.~Hu,
  ``Conformal energy-momentum tensor in curved spacetime: Adiabatic regularization and renormalization,''
  Phys.\ Rev.\ D {\bf 10}, 3905 (1974).
  %doi:10.1103/PhysRevD.10.3905
  %%CITATION = doi:10.1103/PhysRevD.10.3905;%%
  %147 citations counted in INSPIRE as of 26 Sep 2016


%\cite{Bunch:1978gb}
\bibitem{Bunch:1978gb} 
  T.~S.~Bunch,
  ``Calculation of the Renormalized Quantum Stress Tensor by Adiabatic Regularization in Two-Dimensional and Four-Dimensional Robertson-Walker Space-Times,''
  J.\ Phys.\ A {\bf 11}, 603 (1978).
 % doi:10.1088/0305-4470/11/3/021
  %%CITATION = doi:10.1088/0305-4470/11/3/021;%%
  %25 citations counted in INSPIRE as of 26 Sep 2016
  
  %\cite{Bunch:1980vc}
\bibitem{Bunch:1980vc} 
  T.~S.~Bunch,
  ``Adiabatic Regularization For Scalar Fields With Arbitrary Coupling To The Scalar Curvature,''
  J.\ Phys.\ A {\bf 13}, 1297 (1980).
  %doi:10.1088/0305-4470/13/4/022
  %%CITATION = doi:10.1088/0305-4470/13/4/022;%%
  %85 citations counted in INSPIRE as of 26 Sep 2016


%\cite{Haro:2010mx}
\bibitem{Haro:2010mx} 
  J.~Haro,
  ``Topics in Quantum Field Theory in Curved Space,''
  arXiv:1011.4772 [gr-qc].
  %%CITATION = ARXIV:1011.4772;%%
  %6 citations counted in INSPIRE as of 26 Sep 2016


%\cite{Landete:2013lpa}
\bibitem{Landete:2013lpa} 
  A.~Landete, J.~Navarro-Salas and F.~Torrenti,
  ``Adiabatic regularization and particle creation for spin one-half fields,''
  Phys.\ Rev.\ D {\bf 89}, 044030 (2014)
  %doi:10.1103/PhysRevD.89.044030
  [arXiv:1311.4958 [gr-qc]].
  %%CITATION = doi:10.1103/PhysRevD.89.044030;%%
  %17 citations counted in INSPIRE as of 26 Sep 2016


%\cite{delRio:2014cha}
\bibitem{delRio:2014cha} 
  A.~del Rio, J.~Navarro-Salas and F.~Torrenti,
  ``Renormalized stress-energy tensor for spin-1/2 fields in expanding universes,''
  Phys.\ Rev.\ D {\bf 90}, no. 8, 084017 (2014)
  %doi:10.1103/PhysRevD.90.084017
  [arXiv:1407.5058 [gr-qc]].
  %%CITATION = doi:10.1103/PhysRevD.90.084017;%%
  %6 citations counted in INSPIRE as of 26 Sep 2016



%\cite{Capper:1974ic}
\bibitem{Capper:1974ic} 
  D.~M.~Capper and M.~J.~Duff,
  ``Trace anomalies in dimensional regularization,''
  Nuovo Cim.\ A {\bf 23}, 173 (1974).
  %doi:10.1007/BF02748300
  %%CITATION = doi:10.1007/BF02748300;%%
  %220 citations counted in INSPIRE as of 26 Sep 2016


%\cite{Deser:1976yx}
\bibitem{Deser:1976yx} 
  S.~Deser, M.~J.~Duff and C.~J.~Isham,
  ``Nonlocal Conformal Anomalies,''
  Nucl.\ Phys.\ B {\bf 111}, 45 (1976).
  %doi:10.1016/0550-3213(76)90480-6
  %%CITATION = doi:10.1016/0550-3213(76)90480-6;%%
  %271 citations counted in INSPIRE as of 26 Sep 2016
 
 %\cite{Christensen:1978md}
\bibitem{Christensen:1978md} 
  S.~M.~Christensen and M.~J.~Duff,
  ``New Gravitational Index Theorems and Supertheorems,''
  Nucl.\ Phys.\ B {\bf 154}, 301 (1979).
  %doi:10.1016/0550-3213(79)90516-9
  %%CITATION = doi:10.1016/0550-3213(79)90516-9;%%
  %237 citations counted in INSPIRE as of 26 Sep 2016 
  

%\cite{Brown:1977pq}
\bibitem{Brown:1977pq} 
  L.~S.~Brown and J.~P.~Cassidy,
  ``Stress Tensor Trace Anomaly in a Gravitational Metric: General Theory, Maxwell Field,''
  Phys.\ Rev.\ D {\bf 15}, 2810 (1977).
  %doi:10.1103/PhysRevD.15.2810
  %%CITATION = doi:10.1103/PhysRevD.15.2810;%%
  %131 citations counted in INSPIRE as of 26 Sep 2016


%\cite{Duff:1977ay}
\bibitem{Duff:1977ay} 
  M.~J.~Duff,
  ``Observations on Conformal Anomalies,''
  Nucl.\ Phys.\ B {\bf 125}, 334 (1977).
  %doi:10.1016/0550-3213(77)90410-2
  %%CITATION = doi:10.1016/0550-3213(77)90410-2;%%
  %340 citations counted in INSPIRE as of 26 Sep 2016



%\cite{Duff:1993wm}
\bibitem{Duff:1993wm} 
  M.~J.~Duff,
  ``Twenty years of the Weyl anomaly,''
  Class.\ Quant.\ Grav.\  {\bf 11}, 1387 (1994)
  %doi:10.1088/0264-9381/11/6/004
  [hep-th/9308075].
  %%CITATION = doi:10.1088/0264-9381/11/6/004;%%
  %295 citations counted in INSPIRE as of 26 Sep 2016


%\cite{Endo:1984sz}
\bibitem{Endo:1984sz} 
  R.~Endo,
  ``Gauge Dependence of the Gravitational Conformal Anomaly for the Electromagnetic Field,''
  Prog.\ Theor.\ Phys.\  {\bf 71}, 1366 (1984).
  %doi:10.1143/PTP.71.1366
  %%CITATION = doi:10.1143/PTP.71.1366;%%
  %30 citations counted in INSPIRE as of 26 Sep 2016


%\cite{Toms:2014tia}
\bibitem{Toms:2014tia} 
  D.~J.~Toms,
  ``Local momentum space and the vector field,''
  Phys.\ Rev.\ D {\bf 90}, no. 4, 044072 (2014)
  %doi:10.1103/PhysRevD.90.044072
  [arXiv:1408.0636 [hep-th]].
  %%CITATION = doi:10.1103/PhysRevD.90.044072;%%
  %1 citations counted in INSPIRE as of 26 Sep 2016

%c3
%\cite{Deser:1993yx}
\bibitem{Deser:1993yx} 
  S.~Deser and A.~Schwimmer,
  ``Geometric classification of conformal anomalies in arbitrary dimensions,''
  Phys.\ Lett.\ B {\bf 309}, 279 (1993)
  %doi:10.1016/0370-2693(93)90934-A
  [hep-th/9302047].
  %%CITATION = doi:10.1016/0370-2693(93)90934-A;%%
  %243 citations counted in INSPIRE as of 04 Oct 2016

%c2
%\cite{Vieira:2015oka}
\bibitem{Vieira:2015oka} 
  A.~R.~Vieira, J.~C.~C.~Felipe, G.~Gazzola and M.~Sampaio,
  ``One-loop conformal anomaly in an implicit momentum space
  regularization framework,''
  Eur.\ Phys.\ J.\ C {\bf 75}, no. 7, 338 (2015)
  %doi:10.1140/epjc/s10052-015-3561-z
  [arXiv:1505.05319 [hep-th]].
  %%CITATION = doi:10.1140/epjc/s10052-015-3561-z;%%
  %2 citations counted in INSPIRE as of 06 Oct 2016



%\cite{Anous:2014lia}
\bibitem{Anous:2014lia} 
  T.~Anous, D.~Z.~Freedman and A.~Maloney,
  ``de-Sitter Supersymmetry Revisited,''
  JHEP {\bf 1407}, 119 (2014)
  %doi:10.1007/JHEP07(2014)119
  [arXiv:1403.5038 [hep-th]].
  %%CITATION = doi:10.1007/JHEP07(2014)119;%%
  %8 citations counted in INSPIRE as of 26 Sep 2016

%\cite{Chu:2016uwi}
\bibitem{Chu:2016uwi} 
  C.~S.~Chu and D.~Giataganas,
  ``AdS/dS CFT Correspondence,''
  arXiv:1604.05452 [hep-th].
  %%CITATION = ARXIV:1604.05452;%%
  %1 citations counted in INSPIRE as of 26 Sep 2016



%\cite{Brown:1976wc}
\bibitem{Brown:1976wc} 
  L.~S.~Brown,
  ``Stress Tensor Trace Anomaly in a Gravitational Metric: Scalar Fields,''
  Phys.\ Rev.\ D {\bf 15}, 1469 (1977).
  %doi:10.1103/PhysRevD.15.1469
  %%CITATION = doi:10.1103/PhysRevD.15.1469;%%
  %143 citations counted in INSPIRE as of 26 Sep 2016


%\cite{Bunch:1978yq}
\bibitem{Bunch:1978yq} 
  T.~S.~Bunch and P.~C.~W.~Davies,
  ``Quantum Field Theory in de-Sitter Space: Renormalization by Point Splitting,''
  Proc.\ Roy.\ Soc.\ Lond.\ A {\bf 360}, 117 (1978).
  %doi:10.1098/rspa.1978.0060
  %%CITATION = doi:10.1098/rspa.1978.0060;%%
  %701 citations counted in INSPIRE as of 26 Sep 2016


%\cite{Greiner:1996zu}
\bibitem{Greiner:1996zu} 
  W.~Greiner and J.~Reinhardt,
  ``Field quantization,''
  Springer (1996).
  %25 citations counted in INSPIRE as of 26 Sep 2016


%\cite{Frob:2013qsa}
\bibitem{Frob:2013qsa} 
  M.~B.~Frob and A.~Higuchi,
  ``Mode-sum construction of the two-point functions for the
  Stueckelberg vector fields in the Poincare patch of de
  Sitter space,''
  J.\ Math.\ Phys.\  {\bf 55}, 062301 (2014)
  %doi:10.1063/1.4879496
  [arXiv:1305.3421 [gr-qc]].
  %%CITATION = doi:10.1063/1.4879496;%%
  %10 citations counted in INSPIRE as of 26 Sep 2016



%\cite{Bunch:1979uk}
\bibitem{Bunch:1979uk} 
  T.~S.~Bunch and L.~Parker,
  ``Feynman Propagator in Curved Space-Time: A Momentum Space Representation,''
  Phys.\ Rev.\ D {\bf 20}, 2499 (1979).
  %doi:10.1103/PhysRevD.20.2499
  %%CITATION = doi:10.1103/PhysRevD.20.2499;%%
  %192 citations counted in INSPIRE as of 26 Sep 2016
  

%\cite{Belokogne:2015etf}
\bibitem{Belokogne:2015etf} 
  A.~Belokogne and A.~Folacci,
  ``Stueckelberg massive electromagnetism in curved spacetime: Hadamard renormalization of the stress-energy tensor and the Casimir effect,''
  Phys.\ Rev.\ D {\bf 93}, no. 4, 044063 (2016)
  %doi:10.1103/PhysRevD.93.044063
  [arXiv:1512.06326 [gr-qc]].
  %%CITATION = doi:10.1103/PhysRevD.93.044063;%%


\bibitem{cy}
C.S. Chu and Y. Koyama, work in progress. 


%%%%%%%%%%%%%%%%%%%%%%%%%%%%%%%%%%%%%%%%%%%%%%%%%%%%%%%%%%%%%%%%%%%%%%%%%%%%%%%%%%%%%%%%
\end{thebibliography}
\end{document}